\documentclass{aa}  

\usepackage{graphicx}
\usepackage{txfonts}
\usepackage{hyperref}
\hypersetup{colorlinks=true,citecolor=blue}
\usepackage{multirow}
\usepackage{lscape}
\usepackage{supertabular}

\makeatletter
\DeclareRobustCommand{\HII}{%
  \mbox{H\check@mathfonts\fontsize\sf@size\z@\selectfont\ II\ }%
}
\DeclareRobustCommand{\HI}{%
  \mbox{H\check@mathfonts\fontsize\sf@size\z@\selectfont\ I\ }%
}
\DeclareRobustCommand{\Porb}{%
  \mbox{P$_{\rm{orb}}$}%
}

\makeatother

\begin{document}

   \title{The early evolution of young massive clusters}

   \subtitle{II. The kinematic history of NGC\,6618 / M\,17}

   \author{M. Stoop\inst{1}
          \and
          A. Derkink\inst{1}
          \and
          L. Kaper\inst{1}
          \and
          A. de Koter\inst{1,2}
          \and
          C. Rogers\inst{3}
          \and
          M. C. Ram\'{i}rez-Tannus\inst{4}
          \and
          D. Guo\inst{1}
          \and
          N. Azatyan\inst{5}
          }

   \institute{Anton Pannekoek Institute for Astronomy, University of Amsterdam, Science Park 904, 1098 XH Amsterdam, the Netherlands\\
              \email{m.p.stoop@uva.nl}
         \and
             Institute of Astronomy, KU Leuven, Celestijnenlaan 200 D, 3001 Leuven, Belgium
         \and
             Leiden Observatory, Leiden University, P.O. Box 9513, 2300 RA Leiden, The Netherlands
         \and
             Max Planck Institute for Astronomy, K\"{o}nigstuhl 17, D-69117 Heidelberg, Germany
         \and
             Byurakan Astrophysical Observatory, 0213 Aragatsotn Prov., Armenia
             }

   \date{\today}

 
  \abstract
{Characterising the outcome of the star formation process is key to understand and predict the evolution of stellar populations. Especially the fraction of massive stars in young stellar clusters is of importance as they are the dominant sources of both mechanical and radiative feedback, strongly influencing the thermal and dynamical state of their birth environments, and beyond. Their supernovae may trigger the formation of new generations of stars in neighbouring regions. It turns out that a significant fraction of massive stars escape from their parent cluster via dynamical interactions of single stars and/or multiple stellar systems.}
{M\,17 is the nearest giant H~{\sc ii} region hosting a very young and massive cluster: NGC\,6618. Our aim is to identify stars brighter than G $\lesssim$ 21 mag that belong to NGC\,6618, including the (massive) stars that may have escaped since its formation, and to determine the cluster distance and age.}
{The {\it Gaia} DR3 database was used to identify members of NGC\,6618 based on parallax and proper motion within 9$^{\prime}$ from the cluster centre. We searched for nearby stars in a field of 5$^{\circ}$ around the cluster centre that may have originated from the cluster, and we determined their transverse velocity, kinematic age, and impact parameter.} 
{We identified 42 members of NGC\,6618 of which eight have a spectral type of O, with a mean distance of $1675^{+19}_{-18}$~pc and a (transversal) velocity dispersion of about 3 km~s$^{-1}$, and a radial velocity dispersion of $\sim$ 6 km s$^{-1}$. Another ten O stars are associated with NGC\,6618, but they cannot be classified as members due to poor astrometry and/or high extinction. We have also identified six O star runaways. The relative transverse velocity of these runaways ranges from 10 to 70~km~s$^{-1}$ and their kinematic age ranges from about 100 to 750~kyr. Given the already established young age of NGC\,6618 ($\lesssim 1$ Myr), this implies that massive stars are being ejected from the cluster already directly after (or during) the cluster formation process.}
{When constructing the initial mass function, one has to take into account the massive stars that have already escaped from the cluster, that is, about 30\% of the O stars of the original population of NGC\,6618. The trajectories of the O runaways can be traced back to the central 0.2-0.3 pc region of NGC\,6618. The good agreement between the evolutionary and kinematic age of the runaways implies that the latter provides an independent way to estimate (a lower limit to) the age of the cluster.}


   \maketitle
%

\section{Introduction}
\label{sec:introduction}
Massive stars \citep[with initial mass $\gtrsim$ 9 M$_{\odot}$;][]{Poelarends2008} undergo a core-collapse supernova at the end of their life. The formation mechanism(s) of massive stars and the impact of early feedback on their natal environment are generally not well understood. The youngest massive stars are hard to study from an observational perspective. While massive stars are incredibly luminous, they are relatively scarce. Most massive stars are born in young massive clusters that result from the collapse of a giant molecular cloud. The sample of young massive clusters within 2-4 kpc with ages $\lesssim$ 1-2 Myr is small, creating further limitations \citep[see, e.g.,][]{Kuhn2019,MaizApellaniz2022a_villafranca}. In the first few million years, the massive stars may not have had sufficient time to disperse the natal molecular cloud through their ionising photons and stellar winds \citep{Geen2022}, which causes significant extinction that hampers the observations. As the formation timescale for massive stars is relatively short (several $\sim$ 100 kyr), observing the formation process is a challenge.

The outcome of massive star formation is more clear; most of the massive stars end up in binaries or higher order systems \citep{Chini2012,Kiminki2012,Sana2012,Sana2014}. The binary orbital period (\Porb) distribution of massive stars of ages 2-4 Myr reveals that $\gtrsim$ 70\% transfer mass to their companion at some point in their life, with $\sim$ 50\% of the massive binaries having \Porb\ shorter than one month \citep{Sana2012}. \citet{RamirezTannus2021} suggest that the \Porb\ of massive binaries harden over the first 1-2 Myr of their life. This provides potential evidence that massive binaries are not formed with a relatively short \Porb, but rather harden their orbit after star formation. While this trend of massive binaries hardening is relatively clear, the mechanism with which this occurs has so far not been identified.

An important phenomenon in the early life of a young massive cluster is dynamical interaction, where single-binary, binary-binary (or higher-order) systems can exchange angular momentum \citep{Poveda1967,Leonard1988}. In the case of a significantly close passage, two binaries could exchange companions \citep[see e.g.][]{Gualandris2004} or eject stars at a high velocity, so called runaway stars \citep{Blaauw1961}. This is thought to be most important early on as the young massive cluster should be at its densest during or right after the collapse of the star-forming natal cloud \citep{Clarke1992,Oh2016}. Dynamical interactions are thus closely related to the formation mechanism of stellar clusters.

With the new data releases of \textit{Gaia}, accurate astrometry and photometry are available for a large fraction of the brighter stars within 2-3 kpc \citep[G $\lesssim$ 18-21 mag][]{GaiaCollaboration2016,GaiaCollaboration2022,Babusiaux2022}. Young massive clusters and their member stars can now be separated from interloper field stars in greater detail than ever before \citep[see, e.g.,][]{CantatGaudin2018,CastroGinard2022}. For an increasing number of nearby young massive clusters in our Galaxy, runaways have been found that are produced by dynamical interactions; these runaways can be traced back to their birth sites \citep[see, e.g.,][]{Drew2021,MaizApellaniz2022b_bermuda,Stoop2023}. These runaways may convey information about the initial conditions of young massive clusters, such as the cluster radius, stellar density, initial mass function and age \citep{Clarke1992,Oh2016}.

M\,17 is a nearby giant \HII region located in the Sagittarius spiral arm of our Galaxy. The central young massive cluster NGC\,6618 is one of the youngest known in the Galaxy, with most age estimates $\lesssim$ 1-2 Myr \citep{Hanson1997,Hoffmeister2008,Povich2009,RamirezTannus2017}. The distance to NGC\,6618 and M\,17 has been debated in the literature, ranging from $\sim$ 1.3 to 2.1 kpc \citep{Hanson1997,Povich2007,Hoffmeister2008}. The distance estimate to a spatially co-existing maser source G15.03-0.68 results in $d$ = 1.98$^{+0.14}_{-0.12}$ kpc \citep{Xu2011}. More recently, \textit{Gaia} astrometric distance estimates are consistent with a distance of $d$ $\sim$ 1.6-1.7 kpc, implying that the cluster is significantly closer than the maser source \citep{Kuhn2019,MaizApellaniz2022a_villafranca,Kuhn2021}. \citet{Povich2009} hypothesise the presence of an older progenitor cluster north-east of NGC\,6618, which may explain the coinciding emission bubble and several older OB stars. We follow their nomenclature and refer to this progenitor cluster as NGC\,6618\,PG, and to the emission bubble as M\,17\,EB. 

NGC\,6618 is heavily obscured by dust, with $A_{\rm{V}}$ ranging from $\sim$ 5 to 15 mag for sources visible in the optical and/or near-infrared \citep{Povich2009,RamirezTannus2017,RamirezTannus2018}. The central stellar population in NGC\,6618 contains (at least) 15 O stars and more than $\sim$ 100 B stars \citep{Chini1980,Hoffmeister2008}, not accounting for possible runaways located outside the cluster. Massive young stellar objects have also been identified which are still on the pre-main-sequence \citep{RamirezTannus2017}. The radial velocities of the OB stars show a relatively low dispersion compared to other young massive clusters \citep{Sana2017}. A low binary fraction is not what one expects for massive stars, which is why the massive binaries in NGC\,6618 are hypothesised to have large separations \citep{Sana2017}. One of the implications of this is that dynamical interactions could be even more frequent than initially thought. Wide binaries increase the cross section for dynamical interactions, facilitate transfer of angular momentum, and possibly produce runaways. The same dynamical interactions would cause the wide binaries to harden.

We have studied NGC\,6618 and M\,17 with \textit{Gaia} DR3. Section~\ref{sec:method_data_selection} describes the \textit{Gaia} data processing, membership selection, and the spectral classification of massive stars in NGC\,6618\,(PG). Section~\ref{sec:ngc6618_members} describes the results of the membership selection and key parameters of NGC\,6618. In Section~\ref{sec:runaways_find} and \ref{sec:runaway_parameters} we explain how we searched for runaways coming from NGC\,6618, and determined some of their physical parameters. We provide a discussion in Section~\ref{sec:discussion}, and a conclusion and outlook in Section~\ref{sec:conclusion}.


\section{\textit{Gaia} data release 3}
\label{sec:method_data_selection}
We have searched for candidate members of NGC\,6618 in \textit{Gaia} data release 3 (DR3). First, we gathered all \textit{Gaia} sources in a cone-region with a radius of 0.15 deg. This radius corresponds to $\sim$ 4.5 pc at a distance of $\sim$ 1.7 kpc and is sufficiently large to include members up to several half-light radii (see Section~\ref{sec:ngc6618_members}). A larger search radius increases the odds of mistakenly including field stars present around NGC\,6618. This cone-search region was centred on the brightest and most massive system B189 \citep[also known as CEN 1;][]{Chini1980,Bumgardner1992} located in the core of NGC\,6618, with ($l$, $b$) = (15.0565 deg, --0.6884 deg). This yields at first 3108 sources.

To determine the members of NGC\,6618, we applied a set of corrections and filters\footnote{We note that the astrometry and photometry in the full DR3 is unchanged with respect to the early DR3}. In \textit{Gaia} DR3, the correction to the G-band flux and magnitude for sources with 2 and 6-parameter astrometric solutions is already incorporated in the catalogue. We have applied the correction to the parallax to account for the zero-point offset estimated from quasars \citep{Lindegren2021}. To prevent contamination of spurious astrometric solutions, which are typically caused by over-crowding or binaries, we have applied the following filters: first, the renormalised unit weight error (\texttt{ruwe}) should be less than 1.4. If \texttt{ruwe} is larger than 1.4, it likely indicates that the astrometric solution is unreliable \citep{Lindegren2018}. This may exclude binaries and higher order multiples, possibly biasing our membership against multiplicity. Second, the visibility periods used in the astrometric solution (\texttt{visibility\_periods\_used}) should be 10 or more, which could otherwise indicate astrometric or photometric biases \citep{GaiaCollaboration2021}. Third, we discard sources for which the image parameter determination goodness of fit amplitude (\texttt{ipd\_gof\_harmonic\_amplitude}) is larger than 0.15. Fourth, we exclude sources for which more than one peak (\texttt{ipd\_frac\_multi\_peak}) was identified in more than 10\% of the windows used by \textit{Gaia}. The latter two statistics give indications for crowding or binarity if they exceed their respective threshold \citep{GaiaCollaboration2021}. Last, we remove sources for which more than one source identifier was used in the data processing (\texttt{duplicated\_source}), possibly indicating issues in the astrometric solution. After applying these filters, we are left with 2230 sources.

Next, the sources must have had 5 or 6-parameter astrometric solutions, as the proper motion and parallax play a key role in membership selection. Since the distance to NGC\,6618 is estimated to be in the range of 1.3 to 2.1 kpc, we considered stars with \texttt{parallax} ($\varpi$) smaller than 1.0 mas to be foreground stars. We also required \texttt{parallax\_error} ($\sigma_{\varpi}$) to be smaller than 0.12 mas ($\varpi$/$\sigma_{\varpi}$ $\gtrsim$ 5 at a distance of 1.7 kpc), which allowed to more easily identify whether a source is a member or a field star. This left a total of 498 sources.

While \textit{Gaia} DR3 now provides non-single-star astrometric models for $\sim$ 800,000 sources, the sample of non-single-stars beyond $\sim$ 1 kpc is limited \citep{GaiaCollaboration2022}. We have not found multiple star systems consistent with being member of NGC\,6618.

\subsection{Membership selection using \textit{Gaia}}
\label{sec:membership_selection}
We separated the member stars which are part of NGC\,6618 from field stars. We have applied \textsc{pyupmask}\footnote{\url{https://github.com/msolpera/pyUPMASK}}: the \textsc{python} port that builds upon the unsupervised membership algorithm \textsc{upmask}. We refer to \citet{KroneMartins2014} and \citet{Pera2021} for a detailed description of the \textsc{upmask} algorithm. The ICRS astrometric data and errors are converted to Galactic coordinates with the transformation given in the \textit{Gaia} documentation \footnote{\url{https://gea.esac.esa.int/archive/documentation/GEDR3/Data_processing/chap_cu3ast/sec_cu3ast_intro/ssec_cu3ast_intro_tansforms.html}}.

We have applied \textsc{pyupmask} to the 5D astrometric space consisting of the two Galactic coordinates $l$ and $b$, the parallax $\varpi$, and proper motion $\mu_{\rm{l}^{*}}$ ($\mu_{\rm{l}^{*}}$ $\equiv$ $\mu_{\rm{l}}$cos$b$) and $\mu_{\rm{b}}$. While radial velocities are now available for nearly 34 million stars in \textit{Gaia} DR3, for only $\sim$ 20\% of our 498 sources the radial velocity has been determined \citep{Katz2022}. We have not included the radial velocity as a parameter in \textsc{pyupmask} to avoid introducing biases between stars with and without radial velocity.

 In the inner loop of \textsc{pyupmask}, we used an average of 25 stars per so-called candidate cluster. Candidate clusters are created with the Gaussian mixture model, which was determined by \citet{Pera2021} to be the best performing clustering method. In the outer loop of \textsc{pyupmask}, we took uncertainties into account by re-sampling the astrometric parameters using their respective Gaussian uncertainties ($\sigma_{\mu_{\rm{l}^{*}}}$, $\sigma_{\mu_{\rm{b}}}$, $\sigma_{\varpi}$), ignoring correlations between these parameters and stars. Uncertainties on $l$ and $b$ are assumed to be negligible compared to those of the proper motion and parallax. We used 10,000 iterations for the outer loop, and applied the Gaussian-uniform mixture model to clean out possible false positives.

\begin{figure}
\centering
\includegraphics[width=0.99\linewidth]{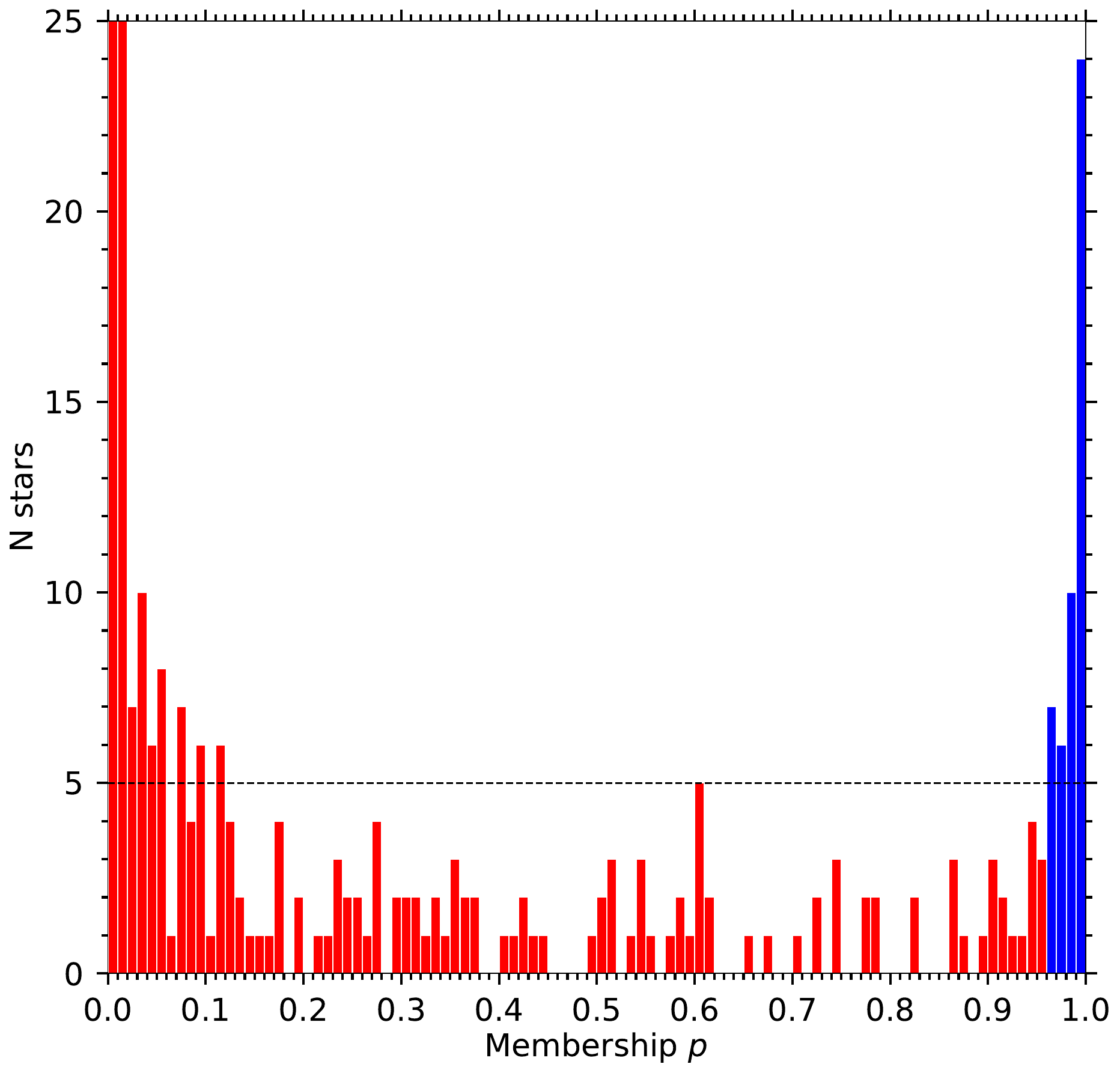}
\caption{Membership probability $p$ distribution for the 498 stars selected in the direction of NGC\,6618 with parallax $<$ 1.0 mas. The field stars have $p$ $<$ 0.96, shown in red, while member stars have $p$ $\geq$ 0.96, shown in blue. The dashed black line (N=5) shows the threshold used for the membership $p$ cut-off between field and member stars. The first and second bin from the left contain 250 and 36 stars, respectively, and are cut off for better visualisation of the other bins.}
\label{fig:member_prob}
\end{figure}

The membership probability $p$ for each star is given by the number of times it is assigned member relative to the total number of iterations. We show the membership $p$ distribution in Figure~\ref{fig:member_prob}, binned in steps of 0.01. We note that the first two bins between $p$ $\in$ [0.00, 0.01] and [0.01, 0.02] contain 250 and 36 members respectively, but have been cut-off for visual clarity. The two distinct peaks in the distribution towards 0.0 and 1.0 show that there is a clear difference between field stars and members. Nevertheless, the field stars heavily outnumber the member stars. Between $p$ of $\sim$ 0.2 and 0.8 there are 59 stars for which their membership is less certain. This is a low level of `noise' of on average one star per bin which should be taken into account when deciding a cut-off membership $p$, to separate the member stars from the field stars. We have decided to constrain this cut-off by having a minimum of five times this `noise level' at five stars per bin (black dashed line in Figure~\ref{fig:member_prob}). This resulted in a minimum membership $p$ of 0.96 and yields 47 member stars. The field and member stars are shown in red and blue, respectively.

As a last step, we removed five stars which deviate by more than 3 standard deviations from the mean for $\mu_{\rm{l}^{*}}$, $\mu_{\rm{b}}$, $\varpi$, ending up with 42 members. We list the members in Table~\ref{tab:members_all}, sorted by their $K_{\rm{s}}$ magnitude and separate them based on whether the stars have been classified on the basis of spectroscopy or not. We have cross-matched the members with the object identifiers in \citet{Bumgardner1992} and \citet{Chini1980}, and list their \textit{2MASS} object identifier otherwise \citep{Skrutskie2006}. We give a complete table of the members with the \textit{Gaia} astrometry and photometry in Appendix~\ref{sec:appendix_members} to avoid cluttering. Only eight of the 42 members have been spectroscopically classified in the literature or here. One of these, B358, is likely a background post-asymptotic-giant-branch star \citep{Chen2013,RamirezTannus2017} not associated with NGC\,6618. This star is located in the centre, with a parallax consistent within 1$\sigma$ with that of NGC\,6618 (see Section~\ref{sec:ngc6618_members}). The proper motion of B358 is marginally consistent within 3$\sigma$. We have not excluded B358 to maintain consistency in our method. Other than B358, we found no evidence for field-star contamination. The (tentative) spectral types found in the literature all indicate the presence of O and B-type stars.

\begin{table*}[]
\caption{\textit{Gaia} DR3 members of NGC\,6618 with and without spectral classification, sorted by their $K_{\rm{s}}$ magnitude.}
\label{tab:members_all}
\begin{tabular}{l l l l l l l}
\hline
\hline
\noalign{\smallskip}Identifier & \texttt{source\_id} & Right ascension & Declination & $K_{\rm{s}}$ & Spectral type & Ref. \\
- & - & deg & deg & mag & - & - \\
\noalign{\smallskip}\hline
\noalign{\smallskip}B189a & 4097815274778072832 & 275.1246 & -16.1790 & 5.7$^{a}$ & O4 V & 2 \\ 
B111 & 4097815382164899840 & 275.1437 & -16.1700 & 7.5 & O4.5 V & 1 \\
B174 & 4097815279085687808 & 275.1268 & -16.1814 &7.6 & O3-6: V & 2 \\
B358 & 4098003055053984256 & 275.0890 & -16.1665 &7.8 & post-AGB & 1\\
2MASS J18202502-1610261 & 4097815313445442816 & 275.1043 & -16.1739 & 7.9 & - & - \\
B137 & 4097815072927248384 & 275.1378 & -16.1893 & 8.1 &  O3-5: V & 2, 3\\
B164 & 4097815347805170944 & 275.1285 & -16.1688 & 8.8 & O6 V & 1 \\
B181 & 4097815279085690240 & 275.1260 & -16.1764 & 9.0 & O9.7 III & - \\
B289 & 4098003085110461312 & 275.1016 & -16.1454 & 9.2 & O9.7 V & 1 \\
2MASS J18202656-1610033 & 4098003020694235904 & 275.1107 & -16.1676 & 9.5 & - & - \\
2MASS J18202528-1610185 & 4098002986334497664 & 275.1054 & -16.1718 & 9.5 & - & - \\
B197 & 4097815244725944064 & 275.1241 & -16.1937 & 9.7 & O9-B1: V & 2 \\
B215 & 4097809163052250624 & 275.1194 & -16.2032 & 10.0 & B0-1 V & 1 \\
2MASS J18202935-1610559 & 4097815244725950720 & 275.1223 & -16.1822 & 10.1 & - & - \\
B205 & 4097815244725949056 & 275.1215 & -16.1863 & 10.2 & B2: V & 2 \\
2MASS J18202802-1610589 & 4097815244725953280 & 275.1168 & -16.1830 & 10.2 & - & - \\
B253 & 4097815309143033088 & 275.1085 & -16.1846 & 10.3 & B3-5 III & 1 \\
B136 & 4097815382164903040 & 275.1378 & -16.1704 & 10.4 & O9: V & 2 \\
CEN 55 & 4097815244725951488 & 275.1214 & -16.1817 & 10.4 & early B & 2 \\
2MASS J18202791-1611087 & 4097815244725951744 & 275.1163 & -16.1858 & 10.6 & - & - \\
2MASS J18202944-1609394 & 4097815450884394112 & 275.1227 & -16.1609 & 10.6 & - & - \\
B93 & 4097815171704114560 & 275.1484 & -16.1821 & 10.7 & B2: V & 2 \\
B213 & 4098003123773451520 & 275.1195 & -16.1573 & 10.8 & O9-B1: V & - \\
2MASS J18202622-1610158 & 4097815313445441664 & 275.1093 & -16.1711 & 10.8 & - & - \\
2MASS J18202622-1609215 & 4098003020694241920 & 275.1092 & -16.1560 & 10.8 & - & - \\
B272 & 4098003020694241024 & 275.1055 & -16.1610 & 10.9 & O9.5: V & 2 \\
B207 & 4097815343502780544 & 275.1203 & -16.1784 & 11.0 & - & - \\
B110 & 4097815171704112256 & 275.1450 & -16.1816 & 11.1 & - & - \\
B234 & 4097815347805176704 & 275.1145 & -16.1704 & 11.1 & - & - \\
2MASS J18202644-1610135 & 4097815347805179904 & 275.1102 & -16.1704 & 11.1 & - & - \\
B150 & 4097815038567510144 & 275.1327 & -16.1940 & 11.2 & O9-B2: V: & 2 \\
2MASS J18202396-1610035 & 4098002986334501504 & 275.0999 & -16.1676 & 11.2 & - & - \\
B118 & 4097815382164896896 & 275.1420 & -16.1790 & 11.4 & - & - \\
2MASS J18202524-1611123 & 4097809437930173440 & 275.1051 & -16.1867 & 11.4 & - & - \\
2MASS J18202655-1610228 & 4097815313445439872 & 275.1106 & -16.1730 & 11.5 & - & - \\
B86 & 4097815176006461440 & 275.1511 & -16.1820 & 11.6 & - & - \\
B201 & 4097809163052248832 & 275.1225 & -16.2035 & 11.6 & - & - \\
2MASS J18203109-1609220 & 4097815450884392704 & 275.1296 & -16.1561 & 11.8 & - & - \\
B56 & 4097815210366202240 & 275.1593 & -16.1690 & 12.5 & - & - \\
2MASS J18202007-1609557 & 4097815450884392704 & 275.1296 & -16.1561 & 12.5$^{b}$ & - & - \\
2MASS J18203617-1611460 & 4097815004207762944 & 275.1507 & -16.1961 & 13.4$^{b}$ & - & - \\
2MASS J18204388-1611236 & 4097814381430170496 & 275.1828 & -16.1899 & 14.1$^{b}$ & - & - \\
\noalign{\smallskip}\hline
\hline
\end{tabular}
\tablefoot{
\tablefoottext{a}{Combined magnitude of B189a and B189b;}
\tablefoottext{b}{Low quality $K_{\rm{s}}$ photometry.}
}
\tablebib{(1) \citet{RamirezTannus2017}; (2) \citet{Hoffmeister2008}; (3) \citet{Povich2009}}
\end{table*}

\begin{table*}
\centering
\caption{O(B) stars in and around NGC\,6618, with and without spectral classification, sorted by their K$_{\rm{s}}$ magnitude. Membership is indicated by `Y'; non-members by `N', and candidate members by `C'.}
\label{tab:member_Ostars}
\begin{tabular}{l l l l l l l l l l}
\hline
\hline
\noalign{\smallskip}\multirow{2}{*}{Identifier} & \multirow{2}{*}{$\alpha$} & \multirow{2}{*}{$\delta$} & \multirow{2}{*}{K$_{\rm{s}}$} & \multirow{2}{*}{Spectral type} & Projected distance & \multirow{2}{*}{Distance} & \multirow{2}{*}{Ruwe} & \multirow{2}{*}{Member} & \multirow{2}{*}{Ref.} \\
& & & & & from centre & & & & \\
\noalign{\smallskip}- & deg & deg & mag & - & arcmin (pc) & pc & - & - & - \\
\noalign{\smallskip}\hline
\noalign{\smallskip}\multicolumn{10}{c}{Confirmed \textbf{O} stars on basis of spectroscopy, this work} \\
\noalign{\smallskip}\hline
\noalign{\smallskip}BD-164831 & 275.3346 & -15.9865 & 7.5 & O9.7 Ia & 16.8 (8.3) & 2437$^{+146}_{-107}$ & 0.970 & N & - \\
\noalign{\smallskip}B260 & 275.1078 & -16.1423 & 7.8 & O9.5 V & 2.2 (1.1) & - & 8.380 & C & - \\
\noalign{\smallskip}BD-164834 & 275.3659 & -15.9591 & 7.9 & O9.5 II & 19.2 (9.5) & - & 7.072 & C & - \\
\noalign{\smallskip}LS4943 & 275.2547 & -16.0964 & 8.9 & O9.7 V & 9.1 (4.5) & 1594$^{+47}_{-40}$ & 0.869 & N & - \\
\noalign{\smallskip}B181 & 275.1260 & -16.1764 & 9.0 & O9.7 III & 0.3 (0.1) & 1772$^{+237}_{-129}$ & 1.174 & Y & - \\
\noalign{\smallskip}LS4941 & 275.2412 & -15.8968 & 9.5 & O9.7 V & 18.1 (9.0) & 1562$^{+41}_{-36}$ & 0.765 & N & - \\
\noalign{\smallskip}\hline
\noalign{\smallskip}\multicolumn{10}{c}{Confirmed \textbf{O} stars on basis of spectroscopy, literature} \\
\noalign{\smallskip}\hline
\noalign{\smallskip}B189a & 275.1246 & -16.1790 & 5.7 & O4 V & 0.2 (0.1) & - & 1.240 & Y & 1 \\
\noalign{\smallskip}B189b & 275.1242 & -16.1793 & 5.7 & O4 V & 0.2 (0.1) & - & 1.418 & C & 1 \\
\noalign{\smallskip}\multirow{2}{*}{BD-164826} & \multirow{2}{*}{275.2593} & \multirow{2}{*}{-16.0169} & \multirow{2}{*}{7.3} & O5 V((f))z & \multirow{2}{*}{12.4 (6.1)} & \multirow{2}{*}{1741$^{+58}_{-48}$} & \multirow{2}{*}{1.019} & \multirow{2}{*}{N} & \multirow{2}{*}{2} \\
& & & & + O9/B0 V & & & & & \\
\noalign{\smallskip}B111 & 275.1437 & -16.1700 & 7.5 & O4.5 V & 1.4 (0.7) & 1817$^{+76}_{-60}$ & 1.109 & Y & 3 \\
\noalign{\smallskip}B0 & 275.1143 & -16.2253 & 7.5 & O6.5 V((f))z & 3.0 (1.5) & 1597$^{+48}_{-40}$ & 1.016 & N & 4 \\
\noalign{\smallskip}B98 & 275.1474 & -16.1802 & 7.6 & O9.5 V & 1.5 (0.8) & - & 19.488 & C & 4 \\
\noalign{\smallskip}B164 & 275.1284 & -16.1688 & 8.8 & O6 Vz & 0.6 (0.3) & 1637$^{+70}_{-56}$ & 1.123 & Y & 3 \\
\noalign{\smallskip}B311 & 275.0946 & -16.1428 & 8.9 & O8.5 Vz & 2.5 (1.3) & 1391$^{+48}_{-40}$ & 1.032 & N & 3 \\
\noalign{\smallskip}B289 & 275.1016 & -16.1454 & 9.2 & O9.7 V & 2.2 (1.1) & 1694$^{+105}_{-76}$ & 1.177 & Y & 3 \\
\noalign{\smallskip}\hline
\noalign{\smallskip}\multicolumn{10}{c}{Candidate \textbf{O} stars on basis of photometry, literature or this work} \\
\noalign{\smallskip}\hline
\noalign{\smallskip}SLS373 & 275.1495 & -16.2620 & 6.8 & O3-6: V & 5.4 (2.7) & 1735$^{+129}_{-89}$ & 0.884 & N & 5 \\
\noalign{\smallskip}B174 & 275.1268 & -16.1814 & 7.6 & O3-6: V & 0.4 (0.2) & 1995$^{+572}_{-236}$ & 1.147 & Y & 1 \\
\noalign{\smallskip}SLS17 & 275.0125 & -16.0352 & 7.8 & O6-9: V & 10.5 (5.2) & 1686$^{+73}_{-58}$ & 1.042 & N & 5 \\
\noalign{\smallskip}2MASS J18182392 & \multirow{2}{*}{274.5996} & \multirow{2}{*}{-17.3644} & \multirow{2}{*}{7.9} & \multirow{2}{*}{O7-B0: V} & \multirow{2}{*}{77 (38)} & \multirow{2}{*}{1755$^{+76}_{-60}$} & \multirow{2}{*}{0.867} & \multirow{2}{*}{N} & \multirow{2}{*}{-} \\
-1721517 & & & & & & & & & \\
\noalign{\smallskip}B137 & 275.1378 & -16.1893 & 8.1 & O3-5: V & 1.2 (0.6) & 1803$^{+338}_{-155}$ & 1.063 & Y & 1, 5 \\
\noalign{\smallskip}HCS6500 & 275.1111 & -16.1191 & 8.6 & O5-7: V & 3.5 (1.7) & 1648$^{+262}_{-131}$ & 1.195 & N & 1, 5 \\
\noalign{\smallskip}B197 & 275.1241 & -16.1937 & 9.7 & O9-B1: V & 1.1 (0.5) & 1684$^{+197}_{-114}$ & 1.208 & Y & 1, 5 \\
\noalign{\smallskip}\hline
\noalign{\smallskip}\multicolumn{10}{c}{Confirmed \textbf{B} stars on basis of spectroscopy, this work} \\
\noalign{\smallskip}\hline
\noalign{\smallskip}BD-164822 & 275.1987 & -16.0267 & 8.2 & B2.5 II & 10.0 (5.0) & 1593$^{+49}_{-41}$ & 0.999 & N & - \\
\noalign{\smallskip}\multirow{2}{*}{BD-154928} & \multirow{2}{*}{275.0575} & \multirow{2}{*}{-15.6642} & \multirow{2}{*}{8.3} & B0.5 V & \multirow{2}{*}{30.9 (15.3)} & \multirow{2}{*}{1676$^{+63}_{-52}$} & \multirow{2}{*}{1.114} & \multirow{2}{*}{N} & \multirow{2}{*}{-} \\
& & & & + B1.5 V & & & & & \\
\noalign{\smallskip}\multirow{2}{*}{LS4972} & \multirow{2}{*}{275.5131} & \multirow{2}{*}{-15.7727} & \multirow{2}{*}{8.7} & B1 V & \multirow{2}{*}{33.1 (16.4)} & \multirow{2}{*}{1591$^{+50}_{-42}$} & \multirow{2}{*}{0.750} & \multirow{2}{*}{N} & \multirow{2}{*}{-} \\
& & & & + B2 V & & & & & \\
\noalign{\smallskip}\multirow{2}{*}{BD-164832} & \multirow{2}{*}{275.3580} & \multirow{2}{*}{-16.0217} & \multirow{2}{*}{8.9} & B0 V & \multirow{2}{*}{16.5 (8.2)} & \multirow{2}{*}{1630$^{+55}_{-46}$} & \multirow{2}{*}{0.846} & \multirow{2}{*}{N} & \multirow{2}{*}{-} \\
& & & & + B1 V & & & & & \\
\noalign{\smallskip}\hline
\hline
\end{tabular}
\tablebib{(1) \citet{Hoffmeister2008}; (2) \citet{MaizApellaniz2019}; (3) \citet{RamirezTannus2017}; (4) \citet{MaizApellaniz2022a_villafranca}; (5) \citet{Povich2009}; (6) \citet{Povich2017}}
\end{table*}

\subsection{Spectroscopic observations}
Only for the bright end of the stellar population in the M\,17 region spectroscopy is available. Spectra are important to asses the spectral type and determine the radial velocity of the stars. We have obtained spectra for 10 bright candidate OB stars in the M\,17 region with either the High Resolution Spectrograph (HRS) mounted on the Southern African Large Telescope (SALT) in Sutherland, the Intermediate-dispersion Spectrograph and Imaging System (ISIS) mounted on the William Herschel Telescope (WHT) at La Palma, and the medium resolution spectrograph X-shooter on the Very Large Telescope (VLT) at Paranal. We observed OB stars with relatively low extinction (WHT), three were part of a monitoring campaign of massive young stellar objects in M\,17 (SALT, programme 2022-1-SCI-002), and a few that are part of a multiplicity study in M\,17 (VLT). We describe and classify the stars in Appendix~\ref{sec:appendix_spectroscopy}; see Table~\ref{tab:member_Ostars}.

\subsection{O stars in and around NGC\,6618}
O stars are the brightest objects in a young stellar population and therefore make up a relatively complete sample. Also, they are potential runaways as the fraction of runaways is claimed to be a strong function of spectral type \citep{deWit2005}. We have compiled information on all 22 confirmed and candidate O stars in and around NGC\,6618 and NGC\,6618\,PG in Table~\ref{tab:member_Ostars}. For 15 of these sources spectra are available that allow to confirm an O-star nature; for six out of these 15 this identification is based on new data presented in Appendix~\ref{sec:appendix_spectroscopy}. For seven of the sources no spectra are available and they are considered O-star candidates on the basis of their K$_{\rm{s}}$ magnitude. Dwarf O stars should have M$_{\rm{K_{s}}}$ $\lesssim$ --3.0 mag \citep{Martins2006,Pecaut2013}. For a distance modulus of 10.6-11.6 ($d$ $\sim$ 1.3 to 2.1 kpc), this yields K$_{\rm{s}}$ $\lesssim$ 8.6 mag for O stars in and around NGC\,6618. However, the central cluster in NGC\,6618 suffers heavily from extinction by dust. Allowing for $\sim$ 10 magnitudes of extinction in the V-band, stars would have $\sim$ 1 magnitude of extinction in the K$_{\rm{s}}$-band. We therefore required candidate O stars to have K$_{\rm{s}}$ $<$ 10.0 mag. This does not exclude missing out on O stars with $A_{\rm{V}}$ $\gtrsim$ 15 mag which could create a bias, but such stars are too faint for \textit{Gaia} anyway. Table~\ref{tab:member_Ostars} provides for each O star or O-star candidate their separation from the centre of NGC\,6618, distance determined from their individual parallax, \texttt{ruwe} and their member classification (true, false, or candidate). In the case that membership could not be determined due to \texttt{ruwe} $>$ 1.4, we have denoted their member classification as `candidate' (C). Several O stars have \texttt{ruwe} $>$ 1.4 so that we can not determine their distance. 

\begin{figure*}
\centering
\centering
\includegraphics[width=0.99\linewidth]{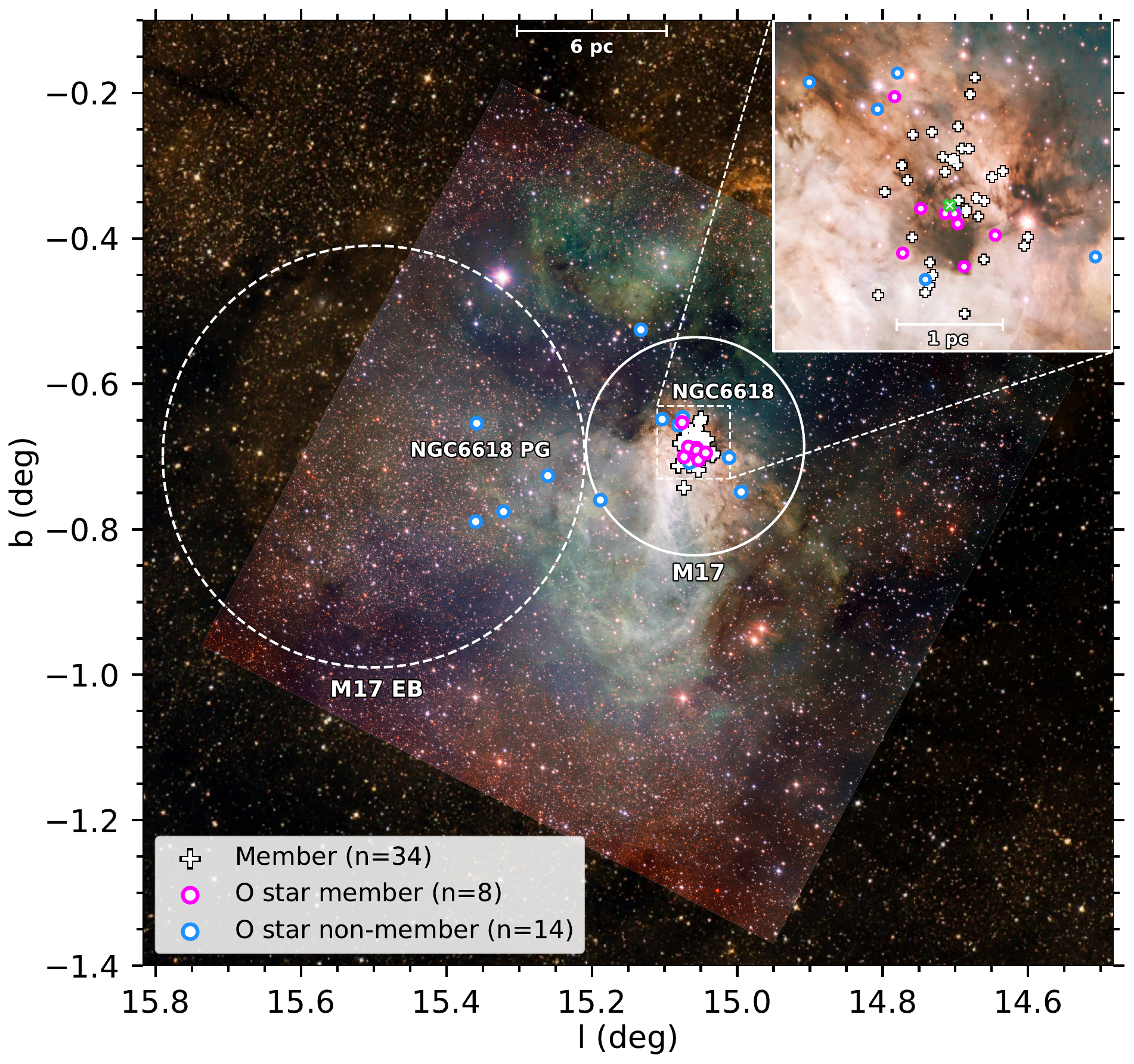}
\caption{Position of all identified O stars or O star candidates in the search field centred on NGC\,6618 and NGC\,6618\,PG, save one. The O star candidate 2MASS\,J18182392-1721517 is located outside of the field shown. Identified member stars are shown with white plusses; those that are O star (candidates) as magenta circles. O star (candidates) that are not identified as member of NGC\,6618 are shown as blue circles. For context, we show the DSS2 B, R and I colour image and the Very Large Telescope Survey Telescope OmegaCAM image (ESO/INAF-VST/OmegaCAM). The inset zooms in on the centre of NGC\,6618. The green cross in this inset indicates the determined cluster centre. With dashed circles the location of the emission nebulae and open cluster(s) are shown. The bars denoting physical size correspond to $\sim 0.2$ deg (6 pc) and $\sim 0.03$ (1 pc).}
\label{fig:l_b_members_ostars}
\end{figure*}

Of the 22 confirmed and candidate O stars, 15 are located near the centre of NGC\,6618 (i.e., they are inside the solid circle in Fig.~\ref{fig:l_b_members_ostars}). The seven that are located further away are not included in the membership search region. Four sources near the cluster centre have \texttt{ruwe} $>$ 1.4 and are also excluded from the membership search. Of the remaining 11 confirmed and candidate O stars, eight are identified as member. The remaining three (with names B0, B189b, and B311) have deviating parallax or proper motion. Still, these three stars are likely part of NGC\,6618 considering their extinction properties and proximity to the cluster centre.

The brightest and most massive system B189 is separated into the a and b component since \textit{Gaia} resolves both. Both B189a and B189b are likely spectroscopic binaries \citep{Hoffmeister2008}, and interferometry also shows that both components could be binaries or triples \citep{Bordier2022}. We have not determined the distance to B189a and B189b as their astrometric solutions are likely spurious.

\begin{figure*}
\centering
\begin{minipage}{.49\textwidth}
\centering
\includegraphics[width=0.95\linewidth]{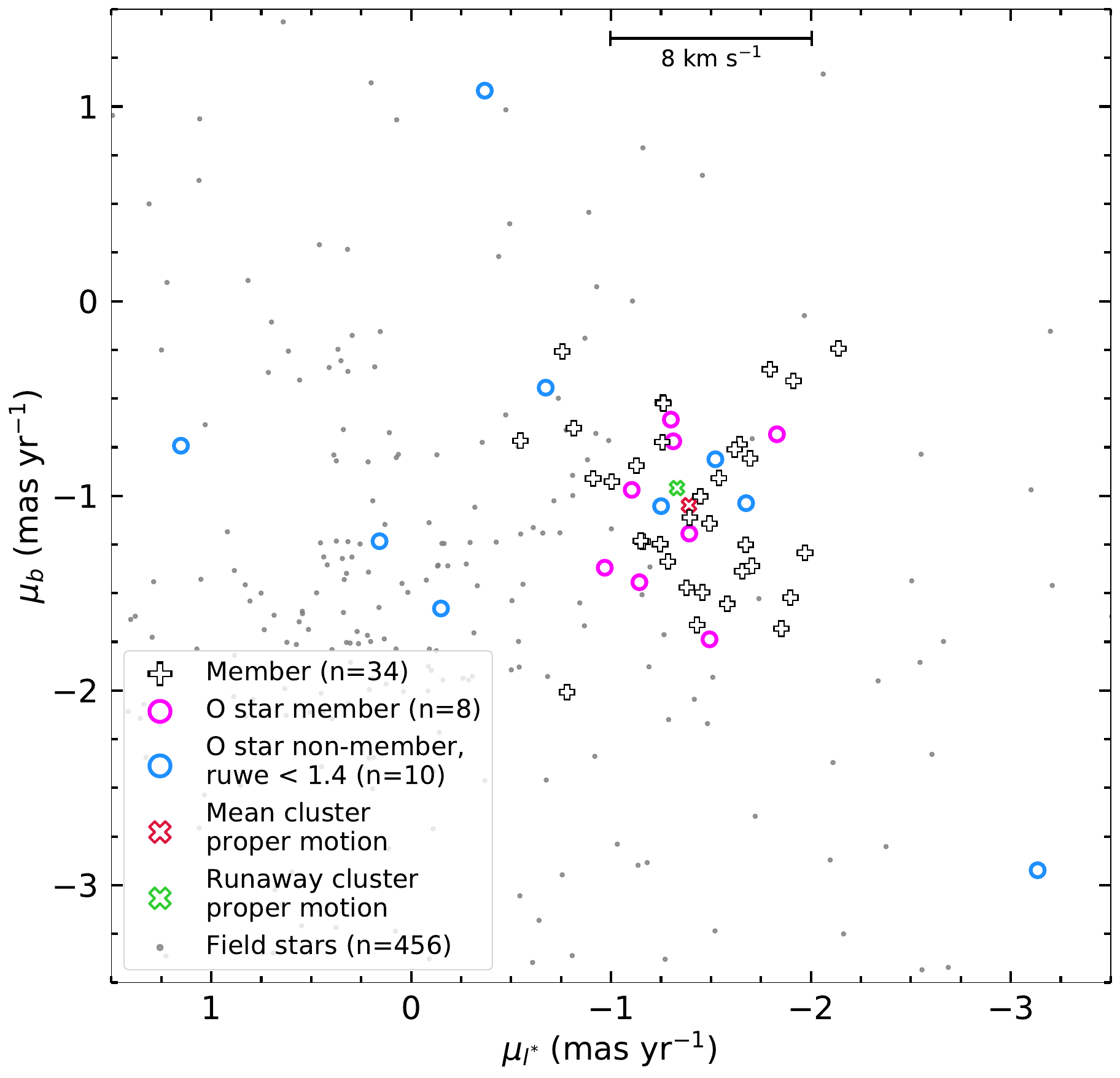}
\end{minipage}
\begin{minipage}{.49\textwidth}
\centering
\includegraphics[width=0.95\linewidth]{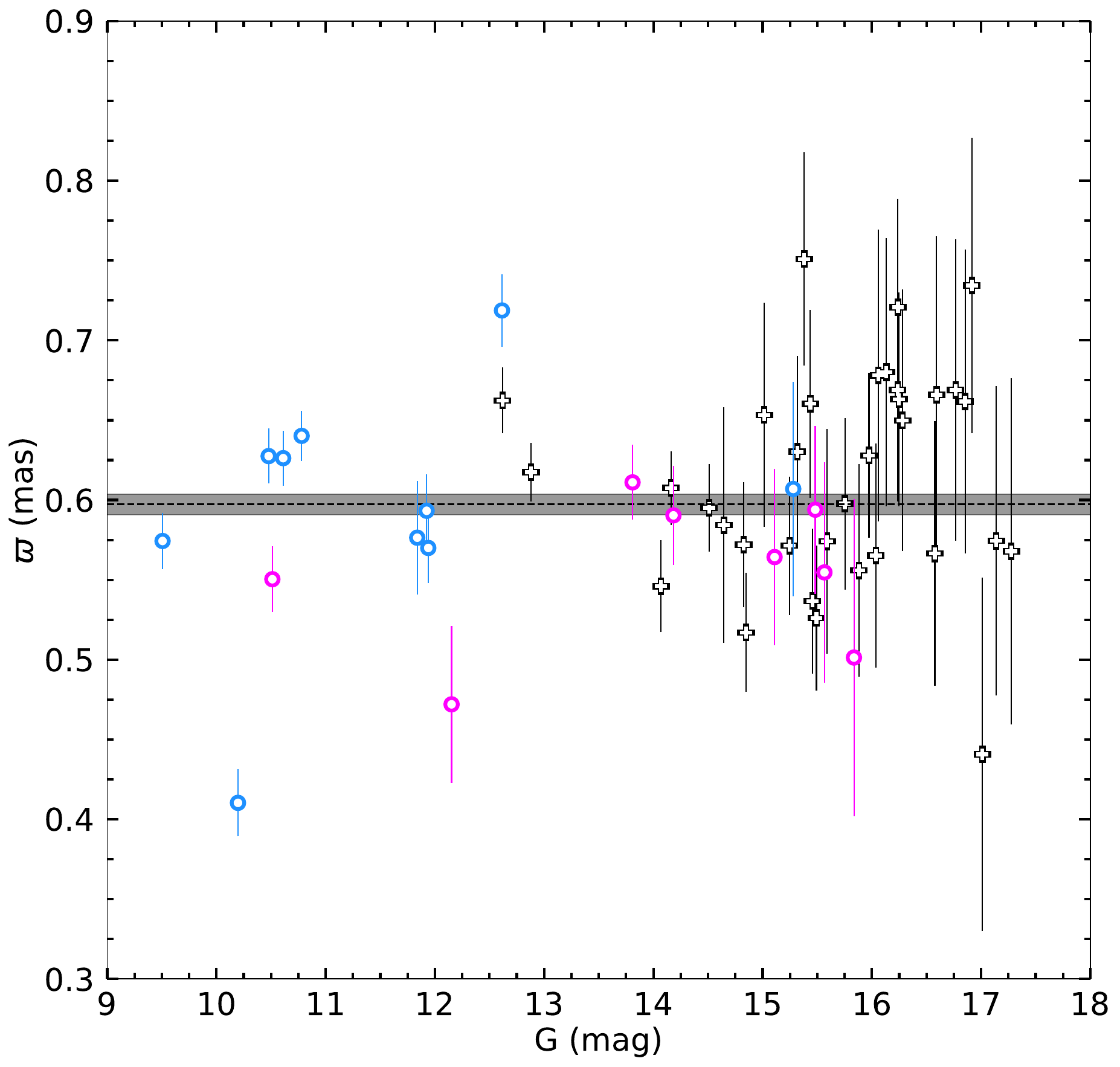}
\end{minipage}
\caption{Proper motion and parallax for members, all identified O-type stars, and field stars in the search field centred on NGC\,6618. The members and O-type stars are coloured and marked similarly as in Figure~\ref{fig:l_b_members_ostars}. O-type stars with \texttt{ruwe} $>$ 1.4 are not shown in both panels. \textit{Left}: Proper motion distribution in the Galactic coordinate frame. We show for context the transverse velocity difference equal to $\sim$ 1 mas yr$^{-1}$ proper motion difference at the distance of NGC\,6618 (1674 pc). The cluster proper motion determined with the mean of the members and from the runaways (see Section~\ref{sec:runaways_find}) are shown with the green and red cross respectively. \textit{Right}: Parallax distribution as a function of the G-magnitude. The best-fit parallax and 1$\sigma$ uncertainty of NGC\,6618 determined from the members are shown with the grey dashed line and bar respectively.}
\label{fig:pm_parallax_members}
\end{figure*}

\section{Members of NGC\,6618}
\label{sec:ngc6618_members}
Studies of membership in Galactic young massive clusters using \textit{Gaia} typically identify $\sim 50$ to 500 members \citep{CantatGaudin2018,Kuhn2019,MaizApellaniz2022a_villafranca}. The exact number depends on several factors, including cluster distance, the amount of line-of-sight and local extinction, the intrinsic number and spatial dispersion of stars in the cluster. We identify 42 members in NGC\,6618, which is at the lower end of what is typical. The detectability of stars in NGC\,6618 suffers heavily from local extinction by dust in the natal molecular cloud. Previous membership analysis of NGC\,6618 from optical and infrared surveys report hundred(s) to thousands of stars belonging to M\,17 \citep{Chini1980,Lada1991,Hanson1997,Hoffmeister2008}. Although we identify far fewer members, the quality of \textit{Gaia} astrometry is such that it allows us to better constrain properties such as the cluster half-light radius, distance, and proper motion. Accurate infrared astrometry that is less affected by extinction, such as the proposed future mission GaiaNIR \citep{Hobbs2016}, could help identify more cluster members.

We show the spatial distribution of the members and all identified O stars in the field in Figure~\ref{fig:l_b_members_ostars}, with the 0.3 deg diameter cone-search region marked by the solid white circle. We have found eight O star members, which we show in magenta. The other 14 O stars, shown in blue, can not be assigned membership due to several reasons. Eleven O stars are located far away from the centre of NGC\,6618, several O stars have \texttt{ruwe} $>$ 1.4 making their astrometry spurious, and several O stars have a significantly deviating proper motion from the cluster average.

The 42 members, shown with the black and white plus symbols, display significant spatial clustering. We show the central 0.1 x 0.1 deg of NGC\,6618 in the inset in the top-right panel in Figure~\ref{fig:l_b_members_ostars}. The members are all concentrated in this inner panel of the cone-search region. Several factors could be responsible for this. First, since NGC\,6618 is young, the members could still be highly concentrated. Second, extreme extinction in the outer regions of NGC\,6618 could obscure our view, causing us to only see the central region. Third, the membership algorithm favours stars in higher stellar density regions, which could also create a bias.

We summarise the astrometric, kinematic and physical properties of NGC\,6618 in Table~\ref{tab:NGC6618params}. To determine the centre of the cluster, we perform Monte Carlo simulations and bootstrapping. We randomly select 80\% of the members (without replacement) and determine for this bootstrapped sample the mean $l$ and $b$. This is repeated $10^{6}$ times to obtain a distribution of means in $l$ and $b$. The centre is determined as the 50$^{\rm{th}}$ percentile, and the 1$\sigma$ uncertainty as the 16$^{\rm{th}}$ and 84$^{\rm{th}}$ percentile of the distribution of means. This results in ($l_{\rm{NGC\,6618}}$, $b_{\rm{NGC\,6618}}$) = (15.058 $\pm$ 0.001 deg, --0.686 $\pm$ 0.001 deg). We show the position of the centre with the green cross in the inset in the top-right panel in Figure~\ref{fig:l_b_members_ostars}, close to the O4 V + O4 V central multiple system B189. That B189 (CEN 1) is exactly in the centre of NGC\,6618 is therefore a good assumption. \citet{Yanza2022} determine the centre from compact radio sources in NGC\,6618. They find ($l$, $b$) = (15.050 deg, --0.689 deg), which is in reasonable agreement considering the aforementioned biases.

The radius of NGC\,6618 can be defined in different ways. Here we determine the 2D projected half-light radius $r_{\rm{hl,2D}}$. To do this, we weigh each member by its relative $K_{\rm{s}}$-flux, which should be least affected by extinction. This results in $r_{\rm{hl,2D}}$ $\sim$ 0.006 deg. Another way to calculate a radius is to take the standard deviation of the 2D projected radii to the centre of the members, which gives $\sim$ 0.011 deg. At the distance of NGC\,6618 (1674 pc, see below), these two radii translate to $\sim$ 0.18 and 0.32 pc respectively. If we assume a Plummer density distribution, the 3D $r_{\rm{hl}}$ will be given by 1.3$r_{\rm{hl,2D}}$ \citep{Plummer1911}, which is 0.23 and 0.42 pc for our two radius definitions, respectively.

\citet{Yanza2022} determine a 2D radius of 0.014-0.018 deg (0.4-0.7 pc) from the two standard deviations on a 2D ellipsoidal Gaussian. This results in a 3D radius of 0.52-0.91 pc assuming again a Plummer density distribution. Our $r_{\rm{hl}}$ is consistent within a factor of two, a result that could be subject to observational biases due to extinction and membership selection. 

We mention that the compact radio sources in \citet{Yanza2022} are likely low mass stars, while our members have higher masses, are brighter, and have higher quality \textit{Gaia} astrometry and photometry. The difference between our determined radius and that of \citet{Yanza2022} could therefore be due to an intrinsically mass dependent spatial distribution of stars. Massive stars could have preferentially formed in or near the centre while lower mass stars could originate from further out regions. A conservative range for the radius of NGC\,6618 $r_{\rm{NGC\,6618}}$ is therefore between 0.2-1.0 pc.

We show the proper motion of the members and O(B) stars in the left panel in Figure~\ref{fig:pm_parallax_members}, adopting similar markers and colours as before. The grey dots show the 456 determined field stars. A large percentage of the O stars with \texttt{ruwe} $<$ 1.4 can be seen to have a deviating proper motion with respect to the members. At the distance of NGC\,6618 (see below), this would correspond to a transverse velocity difference of $\gtrsim$ 10 km s$^{-1}$. We investigate whether this is due to their runaway nature in Section~\ref{sec:runaways_find}. The proper motion of the members is similar compared to other young massive clusters \citep[1--3 km s$^{-1}$;][]{Kuhn2019}. A precise determination of the cluster proper motion is referred to Section~\ref{sec:runaways_find}.

The proper motion dispersion of NGC\,6618 in the $l$ and $b$ direction is $\sigma_{\mu_{l^{*}},\rm{NGC\,6618}}$ = 0.26 mas yr$^{-1}$ and $\sigma_{\mu_{b},\rm{NGC\,6618}}$ = 0.44 mas yr$^{-1}$, respectively. At the distance of NGC\,6618, this translates into $\sigma_{l,\rm{NGC\,6618}}$ = 2.1 km s$^{-1}$ and $\sigma_{b,\rm{NGC\,6618}}$ = 3.5 km s$^{-1}$, respectively. The radial velocity dispersion $\sigma_{\rm{R}}$ of NGC\,6618 is 5.5 $\pm$ 0.5 km s$^{-1}$ \citep{Sana2017,RamirezTannus2021} and differs by a factor of about two with $\sigma_{l}$ and $\sigma_{b}$. A larger $\sigma_{\rm{R}}$ may indicate the presence of binaries, which contribute to the radial velocity dispersion due to orbital motion. Considering that the three velocity dispersions differ by a factor of about two, it is unclear whether these are truly different or are introduced by small number statistics \citep[42 members considered here; 12 stars in][]{RamirezTannus2017}.

We show the parallax of the members and O stars against their G-magnitude in the right panel of Figure~\ref{fig:pm_parallax_members}, again adopting similar markers and colours as before. We determine the distance to NGC\,6618 by setting up a log-likelihood function (without priors) similar to \citet{CantatGaudin2018} and \citet{BailerJones2021} with
\begin{align}
    P(d\ |\ \varpi, \sigma_{\varpi}) &= \prod_{i=1} P(\varpi_{i}\ |\ d, \sigma_{\varpi_{i}}) \nonumber\\&= \prod_{i=1}\frac{1}{\sqrt{2\pi\sigma_{\varpi_{i}}^{2}}}\mathrm{exp}\left(-\frac{(\varpi_{i} - \frac{1}{d})^{2}}{2\sigma_{\varpi_{i}}^{2}}\right),
\end{align}
where $P$ is the unnormalised probability distribution for a given $\varpi$ and assumed Gaussian uncertainty $\sigma_{\varpi}$. The best-fit distance is given by the mode and the 1$\sigma$ uncertainties by the 16$^{\rm{th}}$ and 84$^{\rm{th}}$ percentile of the probability distribution, respectively. We find a distance $d_{\rm{NGC\,6618}}$ = 1674$^{+19}_{-18}$ pc, with a parallax $\varpi_{\rm{M\,17}}$ = 0.5974 $\pm$ 0.0065 mas. The 1$\sigma$ uncertainty on the parallax range is depicted by the grey bar in Figure~\ref{fig:pm_parallax_members}. Most of the members are consistent within 2$\sigma$ with this distance. The north-east component of the visual binary B189 (G $\sim$ 12 mag) shows a deviating parallax with $\varpi$ = 0.472 $\pm$ 0.049 mas, which could be attributed to a spurious astrometric solution due to its multiple nature (not including this star yields a distance of 1667 pc).

The parallax of most of the non-member O stars is consistent within 2$\sigma$ with the determined parallax of NGC\,6618. The clear two outlier O stars, which deviate more than 5$\sigma$ are BD--16 4831 ($\sim$ 2.4 kpc) and B311 ($\sim$ 1.4 kpc). Since BD--16 4831 is in this paper determined to be an O9.7 Ia, this star is significantly more evolved than any age estimate for NGC\,6618. BD--16 4831 is not located in NGC\,6618 and instead is supposedly one of the main ionising sources in NGC\,6618\,PG \citep{Povich2009}. It is unlikely that BD--16 4831 is associated with NGC\,6618\,(PG) and we assume this star to be a background star. The O8.5 Vz star B311 ($d$ $\sim$ 1.4 kpc) is positioned in the centre of NGC\,6618 and has a radial velocity of 4.2 $\pm$ 0.4 km s$^{-1}$ \citep{RamirezTannus2017}, such that it is unlikely to be a radial runaway. B311 suffers from an extinction $A_{\rm{V}}$ $\sim$ 6 mag. With only 2 magnitudes of foreground $A_{\rm{V}}$, B311 is likely located inside NGC\,6618 \citep{Hoffmeister2008}. One might attribute its deviating distance to a spurious astrometric solution, but we find no evidence for this in the \textit{Gaia} astrometry and goodness-of-fit indicators. More investigation is needed to determine whether B311 is a member or foreground star.

\begin{figure}
\centering
\includegraphics[width=0.95\linewidth]{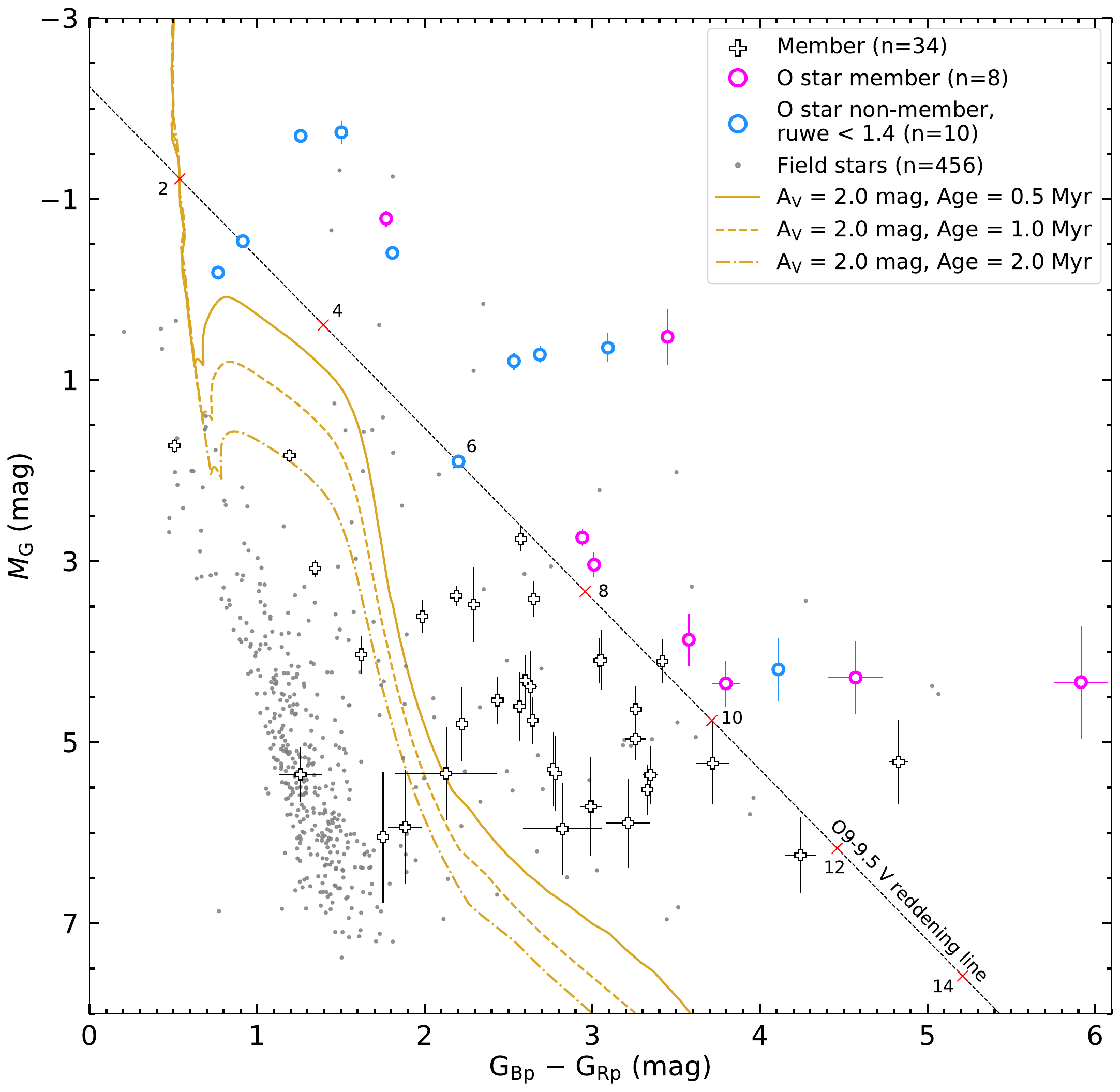}
\caption{Colour - absolute magnitude diagram for members, O-type stars, and field stars in the search field centred on NGC\,6618\,(PG). The members and O stars are coloured and marked similarly as in Figure~\ref{fig:l_b_members_ostars}. Absolute magnitudes are determined from individual parallaxes. Shown are three \textsc{parsec} isochrones with $A_{\rm{V}}$ = 2.0 mag and age of 0.5, 1.0 and 2.0 Myr with the solid, dashed and dot-dashed tracks, respectively. We give the reddening line (R$_{\rm{V}}$ = 3.1) for an O9-9.5 V star with the dashed black line and superimpose on this line the position of 2 to 14 mag of $A_{\rm{V}}$, in steps of 2 mag, with the red crosses.}
\label{fig:camd}
\end{figure}

The colour - absolute magnitude diagram (CAMD) gives us more insight in the age and extinction properties of the members and O stars. We show the CAMD from the \textit{Gaia} photometry in Figure~\ref{fig:camd}, propagating the individual uncertainties on the distance in the $M_{\rm{G}}$ value, and adopting similar colouring and marking as before. The reddening line of an O9-9.5 V star ($R_{\rm{V}}$ = 3.1) is shown with the black dashed line and red crosses superimposed on this line denote $A_{\rm{V}}$ from 2.0 to 14.0 mag in steps of 2.0 mag. The O stars can be seen to follow a similar pattern as the reddening line. The least extincted O stars located in the top-left thus have around $A_{\rm{V}}$ $\sim$ 2.0 mag, while the most extincted O stars have $A_{\rm{V}}$ $\gtrsim$ 10 mag. Figure~\ref{fig:camd} shows that (variable) extinction is a major issue in the nebulous region M\,17. The R$_{\rm{V}}$ also ranges from 3 to 5 within the \HII region \citep{RamirezTannus2018}.

\begin{table}
\centering
\caption{Astrometric, kinematic, and physical parameters of NGC\,6618.}
\label{tab:NGC6618params}
\begin{tabular}{l l l}
\hline
\hline
\noalign{\smallskip}\multicolumn{3}{c}{Equatorial} \\
\noalign{\smallskip}\hline
\noalign{\smallskip}Right Ascension & $\alpha$ & 275.123 $\pm$ 0.001 deg \\
\noalign{\smallskip}Declination & $\delta$ & --16.177 $\pm$ 0.001 deg \\
\noalign{\smallskip}Proper motion\tablefootmark{a} & $\mu_{\alpha^{*}}$ & 0.218 mas yr$^{-1}$ \\
\noalign{\smallskip}Proper motion\tablefootmark{a} & $\mu_{\delta}$ & --1.623 mas yr$^{-1}$ \\
\noalign{\smallskip}$\mu_{\alpha^{*}}$ dispersion & $\sigma_{\alpha}$ & 3.4 km s$^{-1}$ \\
\noalign{\smallskip}$\mu_{\delta}$ dispersion & $\sigma_{\delta}$ & 2.2 km s$^{-1}$ \\
\noalign{\smallskip}\hline
\noalign{\smallskip}\multicolumn{3}{c}{Galactic} \\
\noalign{\smallskip}\hline
\noalign{\smallskip}Galactic longitude & $l$ & 15.058 $\pm$ 0.001 deg \\ 
\noalign{\smallskip}Galactic latitude & $b$ & --0.686 $\pm$ 0.001 deg \\
\noalign{\smallskip}Proper motion\tablefootmark{a} & $\mu_{l^{*}}$ & --1.329 mas yr$^{-1}$ \\
\noalign{\smallskip}Proper motion\tablefootmark{a} & $\mu_{b}$ & --0.957 mas yr$^{-1}$ \\
\noalign{\smallskip}$\mu_{l^{*}}$ dispersion & $\sigma_{l}$ & 2.1 km s$^{-1}$ \\
\noalign{\smallskip}$\mu_{b}$ dispersion & $\sigma_{b}$ & 3.5 km s$^{-1}$ \\
\noalign{\smallskip}\hline
\noalign{\smallskip}\multicolumn{3}{c}{Radial} \\
\noalign{\smallskip}\hline
\noalign{\smallskip}Parallax & $\varpi$ & 0.5974 $\pm$ 0.0065 mas \\
\noalign{\smallskip}Distance & $d$ & 1674$^{+19}_{-18}$ pc \\
\noalign{\smallskip}Radial velocity\tablefootmark{b} & $v_{\rm{R}}$ & 6.4 km s$^{-1}$\\
\noalign{\smallskip}$v_{\rm{R}}$ dispersion$^{(1)}$ & $\sigma_{\rm{R}}$ & 5.5 $\pm$ 0.5 km s$^{-1}$ \\
\noalign{\smallskip}\hline
\noalign{\smallskip}\multicolumn{3}{c}{Physical properties} \\
\noalign{\smallskip}\hline
\noalign{\smallskip}Mass & $M_{\rm{cl}}$ & $\sim$ 5.1 $\times$ $10^{3}$ M$_{\odot}$\\
\noalign{\smallskip}Radius & $r_{\rm{cl}}$ & 0.2-1.0 pc \\
\noalign{\smallskip}Age (runaways) & - & 0.65 $\pm$ 0.25 Myr \\
\noalign{\smallskip}Age (literature) & - & $\lesssim$ 1 Myr \\
\noalign{\smallskip}Visual extinction & $A_{\rm{V}}$ & $>$ 2.0 mag \\
\noalign{\smallskip}Number of O stars & - & $\sim$ 21 \\
\noalign{\smallskip}\hline
\hline\noalign{\smallskip}
\end{tabular}
\tablefoot{
\tablefoottext{a}{Determined with the runaways;}
\tablefoottext{b}{Mean of radial velocities in \citet{RamirezTannus2017}.}
}
\tablebib{(1) \citet{RamirezTannus2021}.}
\end{table}

We include isochrones in Figure~\ref{fig:camd} from the Padova and Trieste Stellar Evolution Code \textsc{parsec} (\texttt{v1.2S}) + \textsc{colibri} (\texttt{S\_37} + \texttt{S\_35} + \texttt{PR16}) models \citep{Bressan2012,Chen2014,Tang2014,Chen2015,Marigo2017,Pastorelli2019,Pastorelli2020}. Shown are three isochrones, each with $A_{\rm{V}}$ = 2.0 mag, but with different ages. Current age estimates for NGC\,6618 are typically$\sim$ 1.0 Myr or younger, with upper limits around 2.0 Myr \citep{Hanson1997,Hoffmeister2008,RamirezTannus2017}. The displayed isochrones have ages of 0.5, 1.0, and 2.0 Myr shown with the solid, dashed, and dash-dotted yellow lines respectively. These isochrones display a vertical main-sequence in the top-left above $M_{\rm{G}}$ $\sim$ 0 mag. Towards fainter and redder magnitudes, they transition to the pre-main-sequence.

Most of the members are to the red of three isochrones. This suggests that these stars have $A_{\rm{V}}$ $\gtrsim$ 2 mag, similar to the O stars. Around seven members are located below G$_{\rm{Bp}}$ -- G$_{\rm{Rp}}$ $\lesssim$ 2 mag and are fainter than the isochrones. We deem these members more likely to be false positive field stars, rather than members with $A_{\rm{V}}$ below 2 mag or ages older than 2.0 Myr. As described in Section~\ref{sec:membership_selection}, we can still expect several false positive field stars.

The members and available isochrones allow us in theory to estimate the age of NGC\,6618 by finding the best-fit isochrone \citep[see e.g.][]{Jorgensen2005}. However, for NGC\,6618 we can clearly see that the spread of the members, caused by strong variable extinction, complicates this method greatly. To solve this, we would need to correct for the variable extinction by obtaining their intrinsic magnitudes and colours through their spectral types. Unfortunately, the spectral types are only available for the brighter and thus more massive stars in NGC\,6618. De-reddening these stars places them on the main-sequence, which gives no information on the age (the main-sequences of the isochrones nearly perfectly overlap in Figure~\ref{fig:camd}).

\section{Dynamically ejected runaways}
\label{sec:runaways_find}
We have searched for runaways originating from NGC\,6618. As NGC\,6618 is estimated to be younger than 2 Myr, runaways are most likely produced by dynamical interactions rather than the supernova mechanism \citep[e.g.][]{vanderMeij2021}. The dynamically ejected runaways are expected to be produced preferentially in the densest region in a cluster \citep{Fujii2011}.

We have searched for runaways within a 5 deg circle centred on NGC\,6618. At a distance of $\sim$ 1.7 kpc, we are able to find a runaway with $v_{\rm{T}}$ = 100 km s$^{-1}$ up to 1.5 Myr ago for this 5 deg radius. These stars are subject to the same filters and corrections as described in Sect.~\ref{sec:method_data_selection}.  ensure that the found runaways come from NGC\,6618 and are not foreground or background stars, the parallax of the runaways should be consistent within 3$\sigma$ with that of NGC\,6618. On top of this, runaways should have a fractional parallax uncertainty $\varpi / \sigma_{\varpi}$ $>$ 10. This stricter requirement on the accuracy in distance than the cut on the parallax to determine membership (see Sect.~\ref{sec:membership_selection}) is to avoid false positives. It also indirectly ensures that the proper motion and thus trace-back of the runaways is accurately known, since these astrometric parameters are correlated.

We have adopted a cut-off magnitude at K$_{\rm{s}}$ = 11 mag, where we have cross-matched the \textit{Gaia} catalogue with the \textit{2MASS} catalogue. This allows us to find not only O-type runaways, but also B-type runaways. The absolute magnitude for each runaway is calculated with their individual parallaxes. For fainter magnitudes, it becomes increasingly more difficult to distinguish between runaways and interloper field stars. The effect of extinction is less by a factor of $\sim$ 10 in the K$_{\rm{s}}$-band than in the G-band, facilitating the discovery of heavily extincted runaways (provided that they have accurate and reliable astrometry).

The runaways should be escaping or have escaped from NGC\,6618. Runaways should therefore be moving away from NGC\,6618 and not towards NGC\,6618, that is, the relative proper motion vector is pointing away from the centre. The velocity dispersion ranges from 2 to 4 km s$^{-1}$, so we require runaways to have $\Delta v_{\rm{T}}$ $>$ 5 km s$^{-1}$ to not pick up on members of NGC\,6618. We note that typical 3D escape velocities are 3-5 km s$^{-1}$ for young massive clusters with masses of $\sim$ $10^{4}$ M$_{\odot}$, while we have only considered 2D velocities. The aforementioned constraints result in a total of 2929 stars.

\begin{figure*}
\centering
\begin{minipage}{.49\textwidth}
\centering
\includegraphics[width=0.95\linewidth]{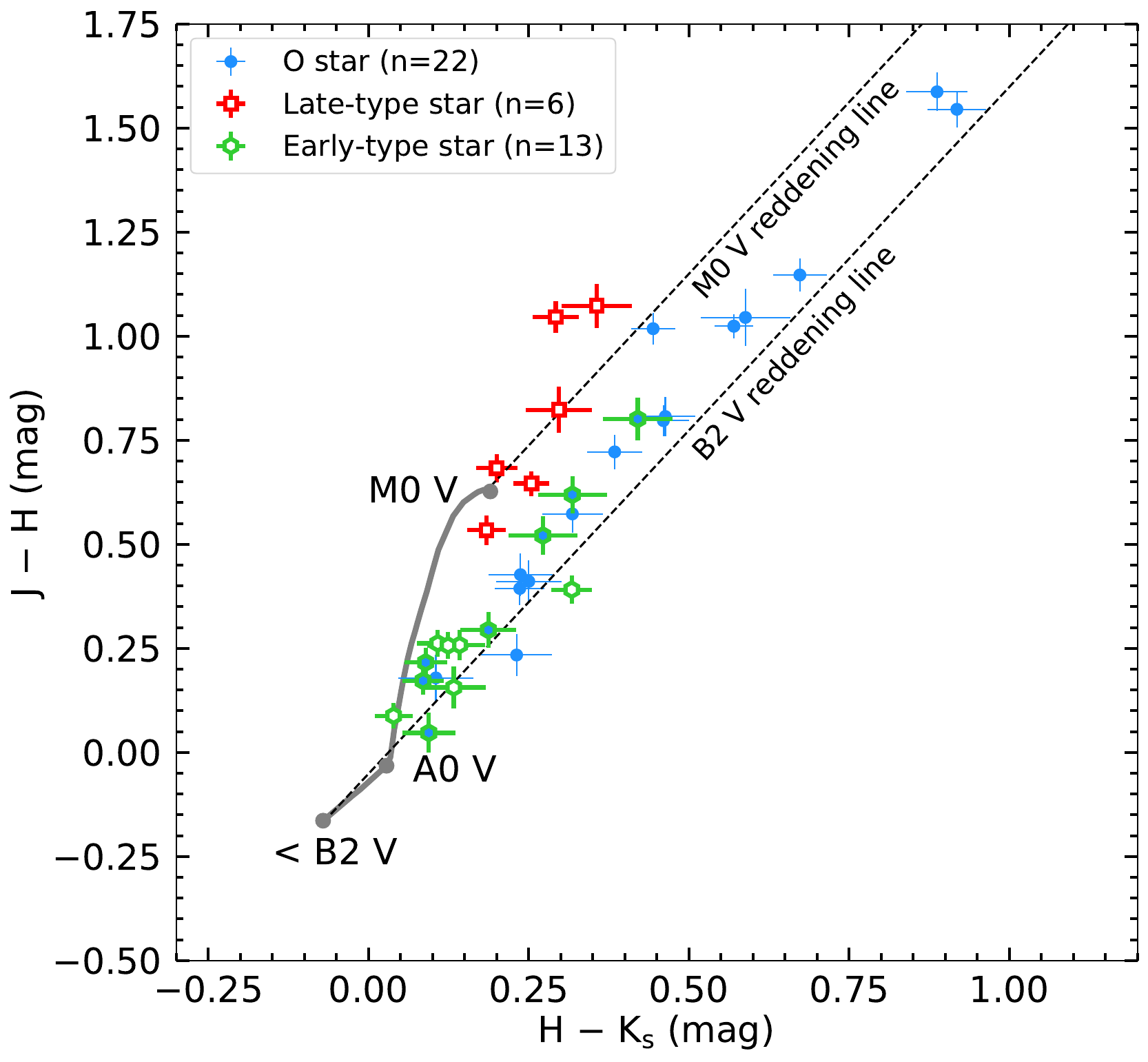}
\end{minipage}
\begin{minipage}{.49\textwidth}
\centering
\includegraphics[width=0.95\linewidth]{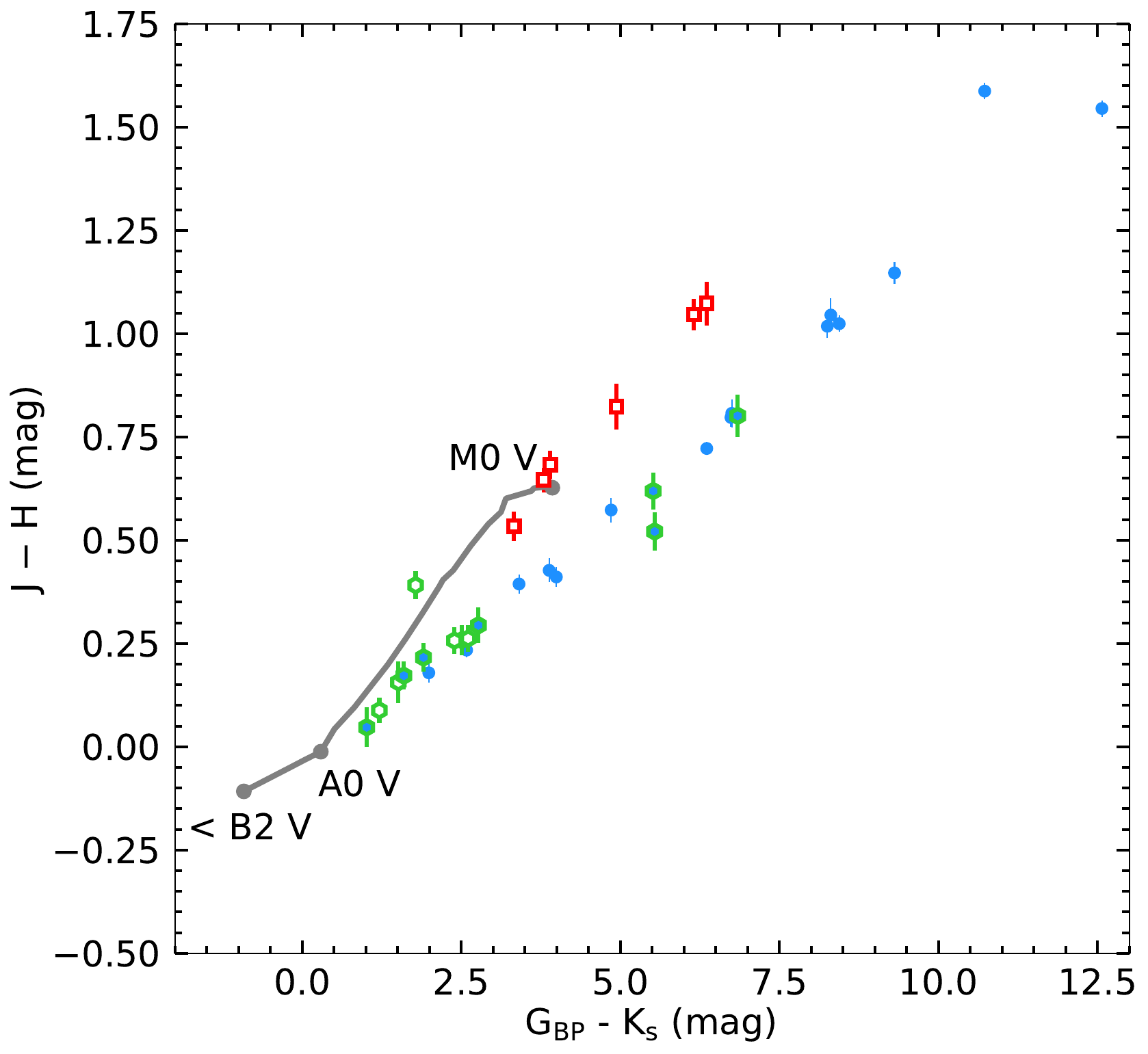}
\end{minipage}
\caption{Colour - colour diagrams for the O stars (blue), late-type stars (red) and early-type runaways (green). \textit{Left}: \textit{2MASS} colour - colour diagram. We indicate the main-sequence with the solid grey line and highlight the position of an M0, A0, and B2 or earlier main sequence star with a grey label and dot. Reddening lines for an M0 V and a B2 V star are plotted assuming $R_{\rm{V}}$ = 3.1. \textit{Right}: \textit{Gaia} and \textit{2MASS} colour - colour diagram.}
\label{fig:runaway_clr_clr}
\end{figure*}

The majority of these stars will move in a `random' direction and only few are actually runaways coming from the cluster. To identify the latter, we have traced back in time the stars to investigate whether they were in NGC\,6618 in the past, using
\begin{align*}
l_{\rm{sep}}(t) &= (l + \frac{t \cdot \mu_{l^{*},\rm{}}}{3.6 \times 10^{6} \cdot \rm{cos} (b)}) \\
&\quad - (l_{\rm{NGC\,6618}} + \frac{t \cdot \mu_{l^{*},\rm{NGC\,6618}}}{3.6 \times 10^{6} \cdot \rm{cos}(b_{\rm{NGC\,6618}})})\ \rm{deg}, \\
b_{\rm{sep}}(t) &= (b + \frac{t \cdot \mu_{b}}{3.6 \times 10^{6}}) \\
&\quad - (b_{\rm{NGC\,6618}} + \frac{t \cdot \mu_{b,\rm{NGC\,6618}}}{3.6 \times 10^{6}})\ \rm{deg},
\end{align*}
where $t$ is the time in years, and $l_{\rm{sep}}$ and $b_{\rm{sep}}$ are the separations between a star and the centre of NGC\,6618. Since NGC\,6618 is located in the Galactic plane, $b$ will likely be between -1 to 0 deg such that the impact of the cos($b$) and cos($b_{\rm{NGC\,6618}}$) factors is relatively small. We also do not account for the gravitational potential of our Galaxy, which is a good approximation up until a few Myr \citep{Hoogerwerf2001}

We expect that true runaways have come from the centre of NGC\,6618. The minimum separation between the runaway candidates and the centre of NGC\,6618 should therefore be relatively small, taking into account the uncertainties in proper motions. With a previously determined $r_{\rm{NGC\,6618}}$ between 0.2-1.0 pc, we require that the runaways come within a conservative separation of 1.0 pc from the centre of NGC\,6618. The runaway candidates are traced back in time for a maximum of 1.5 Myr considering the current age estimates of M\,17 \citep[$\lesssim$ 1 Myr;][]{RamirezTannus2017}. 

While searching for these runaways we found that the minimum separation of the runaway candidates with respect to the cluster centre becomes larger (specifically in $b_{\rm{sep}}$) the longer ago (in time) they were ejected. We attribute this to a marginally different cluster proper motion than what can be determined with the limited sample of 42 members.

To alleviate this issue, we leave both $\mu_{l^{*},\rm{NGC\,6618}}$ and $\mu_{b,\rm{NGC\,6618}}$ as free parameters and adopt an iterative approach. Every iteration, we find runaways satisfying the above conditions for initial cluster proper motions. With the found runaways, we determine new cluster proper motion which minimises the total separation of the found runaways weighted by the number of runaways found. With the new cluster proper motions, we repeat the process of finding runaways, until the cluster proper motion and separation of the found runaways reached a stable minimum. In each iteration, we filter out bright late-type stars (red giants) described below.

The procedure could in theory converge to a non-physical solution, where the separation of only one or two runaways would be minimised. In practice, the first iteration finds 11 runaways and the second (and final) iteration finds 13 runaways with the re-determined cluster proper motion. After this, the runaways, cluster proper motion, and minimum separation were optimised. The re-determined cluster proper motion is $\mu_{l^{*},\rm{NGC\,6618}}$ = --1.33 mas yr$^{-1}$ and $\mu_{b,\rm{NGC\,6618}}$ = --0.96 mas yr$^{-1}$. This differs by $\sim$ 0.07 and 0.1 mas yr$^{-1}$, respectively, from the mean proper motion of the members. At the distance of NGC\,6618, this would correspond to a velocity difference of $\sim$ 1 km s$^{-1}$, or 1 pc Myr$^{-1}$, which is significant compared to the radius of $\lesssim$ 1.0 pc. While we could have increased our adopted $r_{\rm{NGC\,6618}}$, this would cause us to wrongly determine important physical parameters of the runaways such as their impact parameter and kinematic age. A different cluster proper motion causes the (time of) minimum separation to change. We show the proper motion determined with the runaways and with the mean of the members with the green and red cross respectively in the left panel of Figure~\ref{fig:pm_parallax_members}. We adopt the proper motion determined here with the runaways.

The found runaways may contain false positives, that is, interloper field stars that happen to satisfy the conditions listed above. Bright late-type interlopers, such as red giants, are easily identified in colour-colour diagrams, where the effective temperature of stars determines their location. Specifically, we made use of the (J -- H) - (H -- K$_{\rm{s}}$) and the (J -- H) - (G$_{\rm{BP}}$ -- K$_{\rm{s}}$) diagrams, which we show in the left and right panels of Figure~\ref{fig:runaway_clr_clr}, respectively.

\begin{figure*}
\centering
\includegraphics[width=1.0\linewidth]{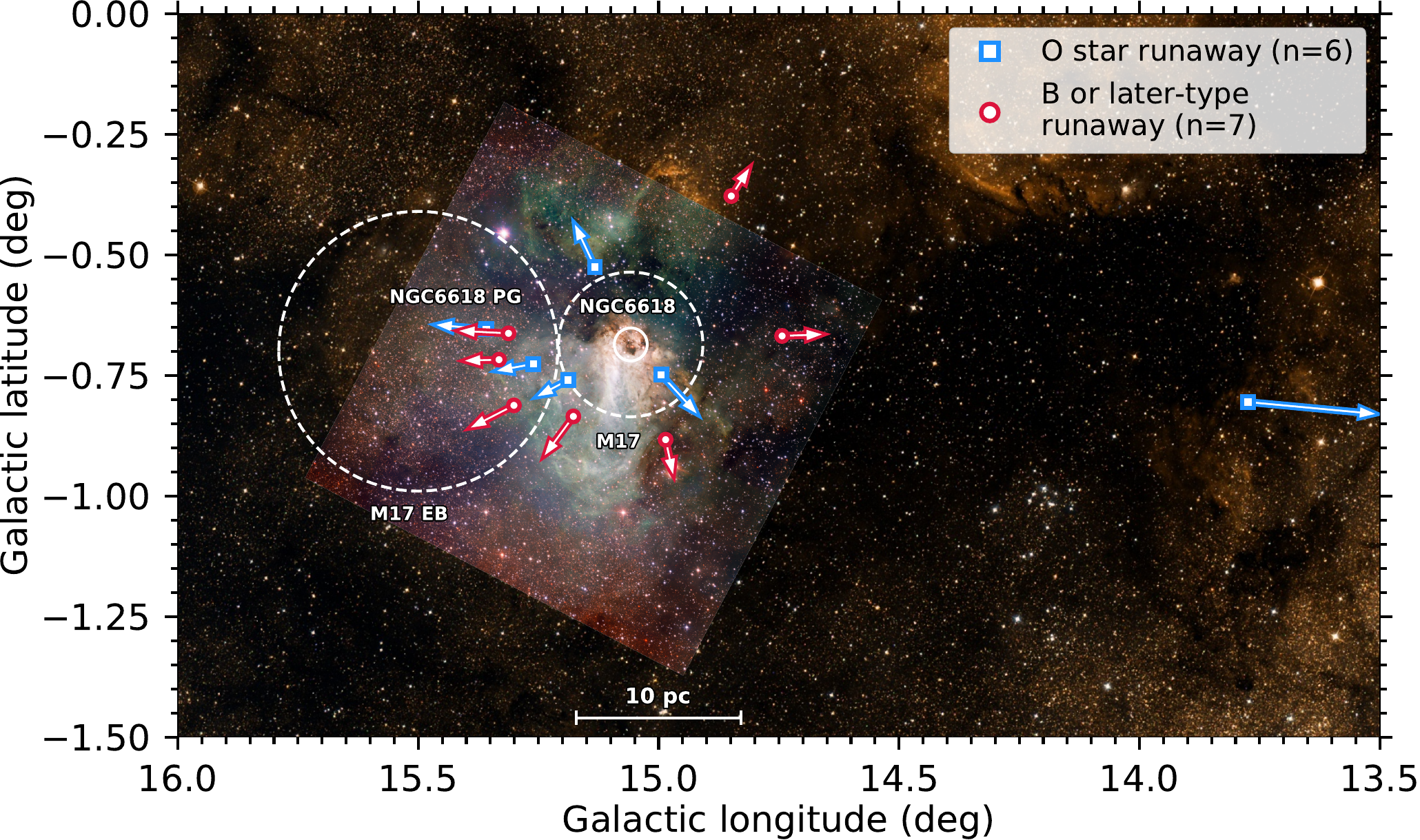}
\caption{Position and motion of the found runaways relative to NGC\,6618, with the DSS2 B, R, and I colour image and the Very Large Telescope Survey Telescope OmegaCAM image (ESO/INAF-VST/OmegaCAM). We show the proper motion of the runaways relative to NGC\,6618 with the arrows, which are proportional to their absolute proper motion. Runaways which are consistent with being O stars in Table~\ref{tab:member_Ostars} are coloured in blue, and coloured red otherwise. We show the position of NGC\,6618\,(PG) and M\,17\,(EB) using white circles. The angular size equivalent to 10 pc is given at the distance of NGC\,6618 (1674 pc).}
\label{fig:runaway_motion}
\end{figure*}

We show the main-sequence track in both panels with the solid grey line, where the early and late-type stars occupy a different position. The O and early-type B stars all have the same intrinsic colours in these diagrams, which we indicate with $<$ B2 V. The black dashed lines show the reddening lines for a B2 V and M0 V star (R$_{\rm{V}}$ = 3.1). Even accounting for reddening, early and late-type stars should be separated from each other. For context, we show the O stars in Table~\ref{tab:member_Ostars} in blue, which mostly follow the B2 V reddening line. The O star B197 is located on the M0 V reddening line in the left panel, which could indicate it is a red giant, since it has only been classified based on photometry. The runaways consistent with being early and late-type stars are shown in green and red, respectively. The six identified late-type stars have colours consistent with late-type stars. 

On top of this, we have investigated the $T_{\rm{eff}}$ and log($g$) estimated with the General Stellar Parameterizer from Photometry (GSP-Phot) in \textit{Gaia} DR3 \citep{BailerJones2010,BailerJones2011,BailerJones2013,Andrea2022}. GSP-Phot infers stellar parameters from the $G_{\rm{BP}}$ and $G_{\rm{RP}}$ spectra. Three of the six identified late-type stars have $T_{\rm{eff}}$ (\texttt{teff\_gspphot} in \textit{Gaia} DR3) in the range of 4500-5500 K, while $T_{\rm{eff}}$ for the early-type stars are almost all not available. Similarly, the log($g$) (\texttt{logg\_gspphot} in \textit{Gaia} DR3) of these three late-type stars are in the range of 1.7-2.7, while those for the early-type stars are again not available. We identify the six late-type stars as false positives and have removed them, leaving us with 13 runaway stars in our final runaway search iteration.

\section{Physical properties of runaways}
\label{sec:runaway_parameters}
We list the spectral and kinematic properties of the runaways in Table~\ref{tab:runaways}. Seven of these runaways are either spectroscopically confirmed or photometrically consistent with being an O star and are also listed in Table~\ref{tab:member_Ostars}. We visualise the position and tangential velocity of the runaways in Figure~\ref{fig:runaway_motion}. The arrows indicate the runaway's motion relative to NGC\,6618, with the length of the arrow proportional to the absolute proper motion. The runaways consistent with being an O star are shown in blue, and the likely B or later-type runaways in red. Six to seven of the thirteen found runaways move in the direction of NGC\,6618\,PG. Several of these runaway stars were identified by \citet{Povich2009} to be part of NGC\,6618\,PG. We discuss this in Section~\ref{sec:disc_ngc6618pg}.

\begin{figure}
\centering
\includegraphics[width=0.95\linewidth]{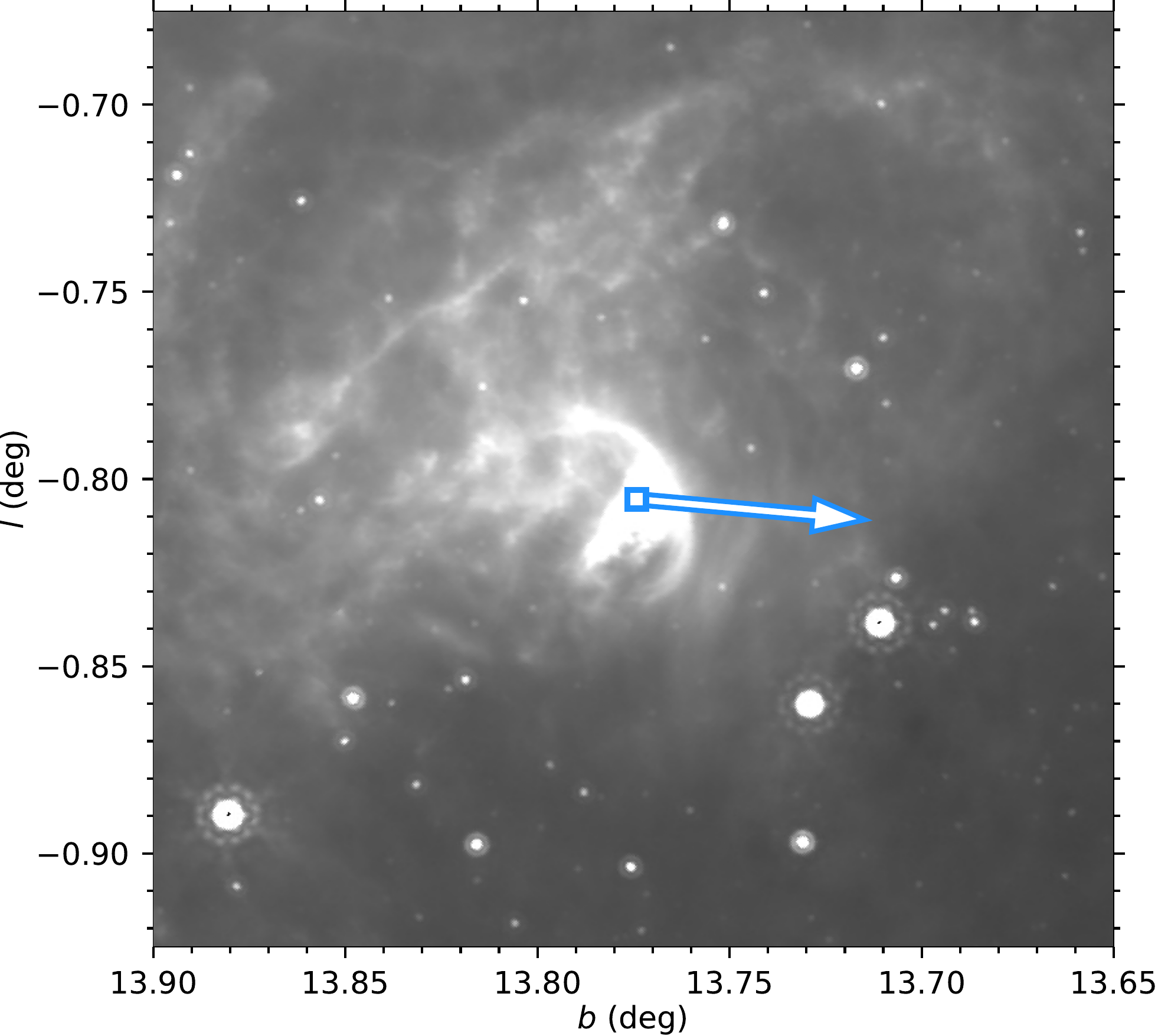}
\caption{\textit{Spitzer} 24 $\mu$m Multi-Band Imaging Photometer grey-scale image of the bowshock around the runaway star 2MASS J18182392-1721517. The blue square shows the position of the star. We show the relative proper motion with respect to NGC\,6618 with the blue arrow.}
\label{fig:bowshock}
\end{figure}

All except one runaway are separated by $\sim$ 0.1-0.4 deg from NGC\,6618, corresponding to $\sim$ 3-12 pc. Only one runaway, \textit{2MASS} J18182392-1721517, is separated by $\sim$ 1.25 deg (35 pc). There are several reports in the literature of excess infrared emission at the position of this runaway, possibly indicating an \HII region or `bubble' \citep{Simpson2012,Su2018}. We show in Figure~\ref{fig:bowshock} the \textit{Spitzer Space Telescope} Multi-Band Imaging Photometer (MIPS) field around \textit{2MASS} J18182392-1721517 \citep[24 $\mu$m;][]{Werner2004,Rieke2004}. We can see that the infrared emission appears to have a bowshock-like shape. The blue arrow indicates the proper motion of this runaway relative to NGC\,6618, from which we can see that the bowshock shape is roughly in the direction of motion. We further discuss the nature of \textit{2MASS} J18182392-1721517 and the bowshock in Section~\ref{sec:disc_bowshock}.

We determine the transverse velocity with respect to NGC\,6618 ($|\Delta v_{\rm{T}}|$) for each runaway from their individual proper motions and parallaxes, propagating their uncertainties. We list $|\Delta v_{\rm{T}}|$ in Table~\ref{tab:runaways}. Almost all runaways have $|\Delta v_{\rm{T}}|$ ranging from $\sim$ 10 to 20 km s$^{-1}$, with only \textit{2MASS} J18182392-1721517 having a significantly higher transverse velocity ($\sim$ 65 km s$^{-1}$). \textit{Gaia} provides radial velocities for two runaways, and no other radial velocities are known in the literature. The stars in NGC\,6618 have a mean radial velocity of $\sim$ 6.4 km s$^{-1}$ \citep{RamirezTannus2017}, indicating that both runaways have radial velocities that deviate from this by more than 3$\sigma$. This could either be due to their runaway nature or to binarity (or both).

We next focus on the kinematic age ($t_{\rm{kin}}$) and the minimum separation between the runaway and cluster centre ($r_{\rm{imp,2D}}$, the projected impact parameter) of the runaways, which require having to deal with several uncertainties. The exact origin of each runaway in NGC\,6618 is unknown. We therefore assume an uncertainty region around the centre of NGC\,6618 with a radius of $\sim$ 1.0 pc, determined to be the upper limit on the cluster radius of NGC\,6618 in Section~\ref{sec:ngc6618_members}. The proper motion of NGC\,6618 has been determined in Section~\ref{sec:runaways_find} and we assume 1$\sigma$ uncertainty of 0.02 mas yr$^{-1}$ in both $\mu_{\rm{l^{*}}}$ and $\mu_{\rm{b}}$. The uncertainty are similar to that of the runaways, which have a 1$\sigma$ uncertainty in $\mu_{\rm{l^{*}}}$ and $\mu_{\rm{b}}$ in the range of 0.01-0.03 mas yr$^{-1}$.

We resort to MC simulations to determine both $t_{\rm{kin}}$ and $r_{\rm{imp,2D}}$ simultaneously. In each iteration, we randomly draw the runaway $\mu_{\rm{l^{*}}}$ and $\mu_{\rm{b}}$ from a 2D Gaussian distribution with mean, standard deviation and covariance matrix equal to the observed value, 1$\sigma$ uncertainty and correlation. The proper motion of NGC\,6618 is also randomly drawn from a 2D Gaussian distribution similar to the runaways, with the uncertainty listed above. In each iteration, we determine the time when the runaway `enters' the uncertainty region around NGC\,6618, is closest to the centre, and `exits' the uncertainty region of NGC\,6618. After 10,000 iterations, we obtain three distributions of time from which we calculate the lower limit, best-fit, and upper limit on $t_{\rm{kin}}$ as the 16$^{\rm{th}}$ percentile, mean, and 84$^{\rm{th}}$ percentile, respectively. We determine the best-fit $r_{\rm{imp,2D}}$ and its uncertainty as the mean and standard deviation on the distribution of closest approach to the centre.

Table~\ref{tab:runaways} lists the determined $t_{\rm{kin}}$ and $r_{\rm{imp,2D}}$ for the runaways. The uncertainty in $t_{\rm{kin}}$ is not symmetrical. Since the uncertainties in the motion of the runaways and cluster increase with time, the positive error will be larger than the negative error when calculating further back in time. We have taken the absolute values to avoid confusion. We note that $r_{\rm{imp,2D}}$ is a projected distance and that we have minimised the total $r_{\rm{imp,2D}}$ in Section~\ref{sec:runaways_find} by having the cluster proper motion as a free parameter.

\begin{figure}
\centering
\includegraphics[width=1.0\linewidth]{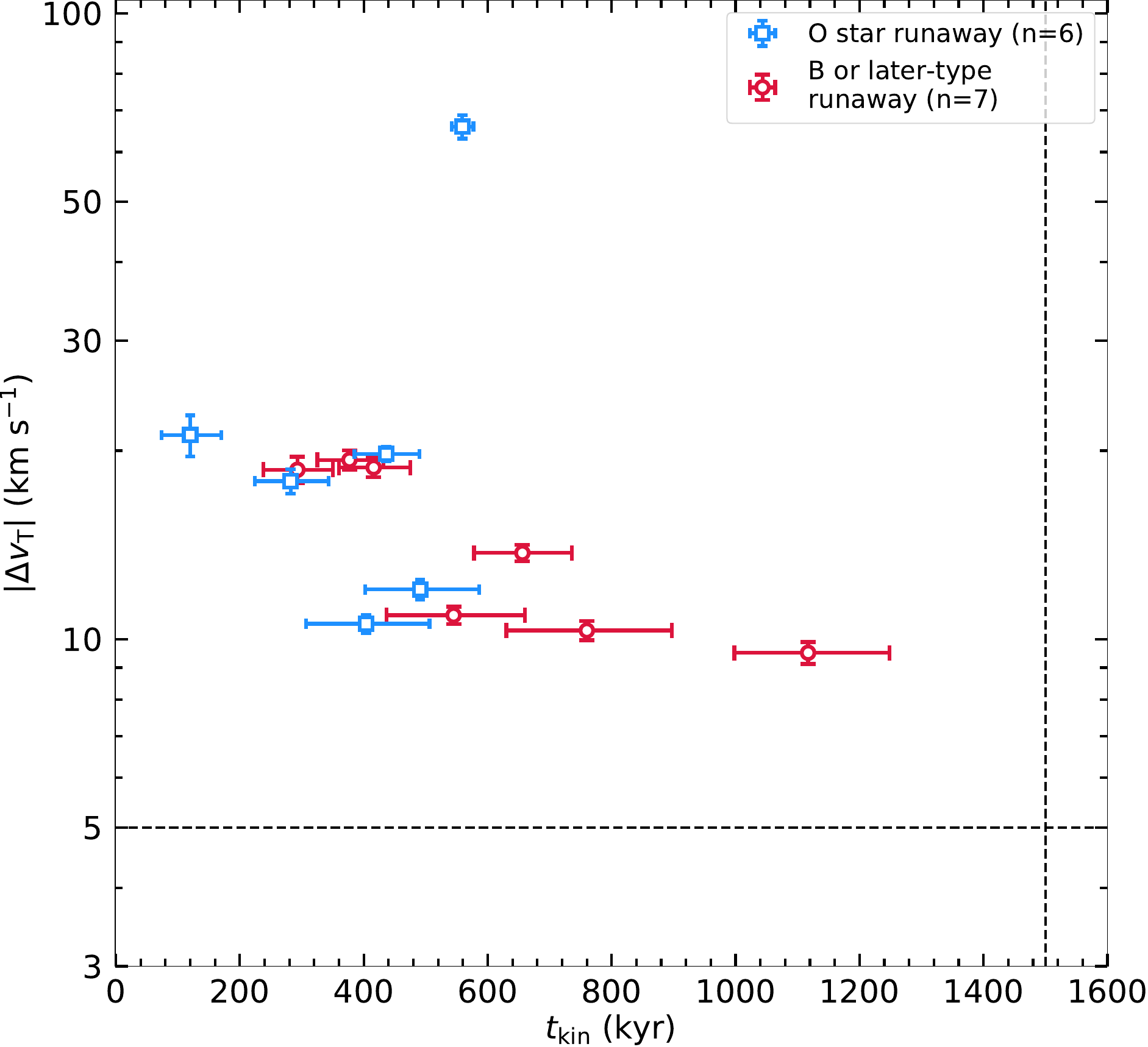}
\caption{Relative transverse velocity ($|\Delta v_{\rm{T}}|$) of the runaways as a function of their kinematic age ($t_{\rm{kin}}$). The identified O stars in Table~\ref{tab:member_Ostars} are shown in blue; red is used for B or later-type stars. The cut-off $|\Delta v_{\rm{T}}|$ and $t_{\rm{kin}}$ used in the runaway search are indicated with the dashed black lines.}
\label{fig:tkin_runaways}
\end{figure}

We show $|\Delta v_{\rm{T}}|$ as a function of $t_{\rm{kin}}$ in Figure~\ref{fig:tkin_runaways}, adopting similar colouring and marking as in Figure~\ref{fig:runaway_motion}. The $t_{\rm{kin}}$ of the runaways ranges from $\sim$ 100 to 1250 kyr, concentrating between 200-700 kyr. The already identified O stars are shown in blue and have $t_{\rm{kin}}$ less than $\sim$ 600 kyr. The latest O star runaway could have been ejected as recently as 70-170 kyr ago. We show the adopted minimum $|\Delta v_{\rm{T}}|$ and maximum $t_{\rm{kin}}$ with the dashed black lines. This shows that the typical runaway $|\Delta v_{\rm{T}}|$ between 10-20 km s$^{-1}$ is not a result of our cut-off. We would have found runaways with $|\Delta v_{\rm{T}}|$ between 5-10 km s$^{-1}$ if they existed. Similarly, we do not find runaways with $t_{\rm{kin}}$ $\gtrsim$ 1200 kyr, making it unlikely that there is a significant runaway population older than this.

\begin{figure}
\centering
\includegraphics[width=1.0\linewidth]{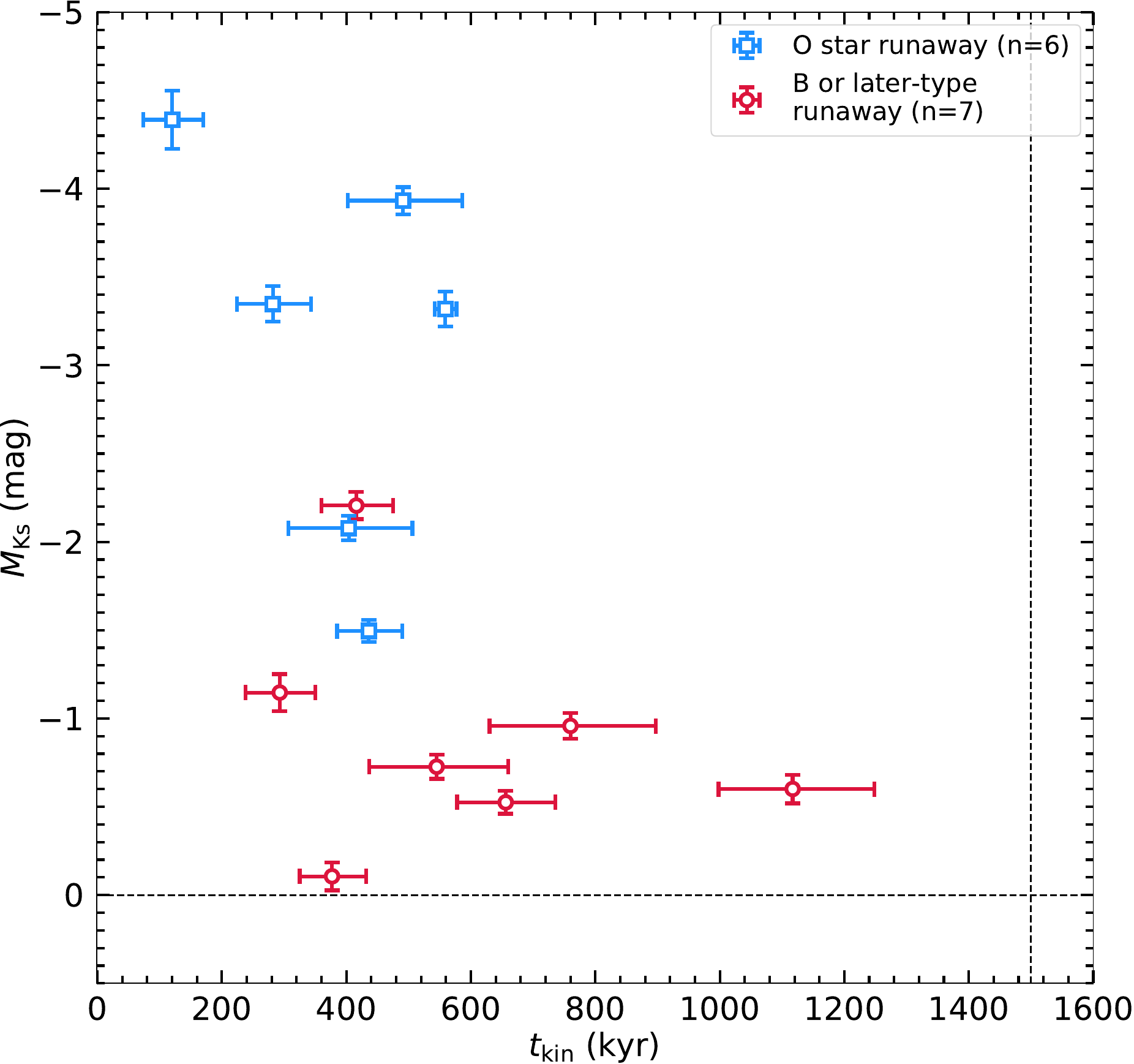}
\caption{Absolute $M_{\rm{K_{s}}}$ magnitude of the runaways as a function of their kinematic age ($t_{\rm{kin}}$), marked and coloured similarly as in Figure~\ref{fig:tkin_runaways}. The cut-off $M_{\rm{K_{s}}}$ and $t_{\rm{kin}}$ used in the runaway search are indicated with the dashed black lines.}
\label{fig:ks_tkin_runaways}
\end{figure}

Figure~\ref{fig:ks_tkin_runaways} shows the absolute $\rm{K}_{\rm{s}}$ magnitudes, determined from their individual parallaxes, adopting similar colours as before. The already spectroscopically confirmed O stars can be seen to be brighter than $M_{\rm{K_{s}}}$ $<$ -1.5 mag, as expected. The spectroscopically identified B0 V + B1 V system is the brightest B or later-type system, with $M_{\rm{K_{s}}}$ $\sim$ -2.2 mag. We are likely missing runaways fainter than $M_{\rm{K_{s}}}$ $\gtrsim$ 0 mag ($\rm{K}_{\rm{s}}$ $>$ 11 mag), indicated with the dashed black line. As previously mentioned, this cut-off is intentional as for fainter magnitudes it becomes increasingly more difficult to distinguish between false positives and true runaways. Figure~\ref{fig:ks_tkin_runaways} shows how the adopted magnitude cut-off significantly impacts the found runaways. If we adopted M$_{\rm{K_{s}}}$ $<$ --1 mag, we would have only found runaways with t$_{\rm{kin}}$ $<$ 600 kyr.

\begin{figure}
\centering
\includegraphics[width=1.0\linewidth]{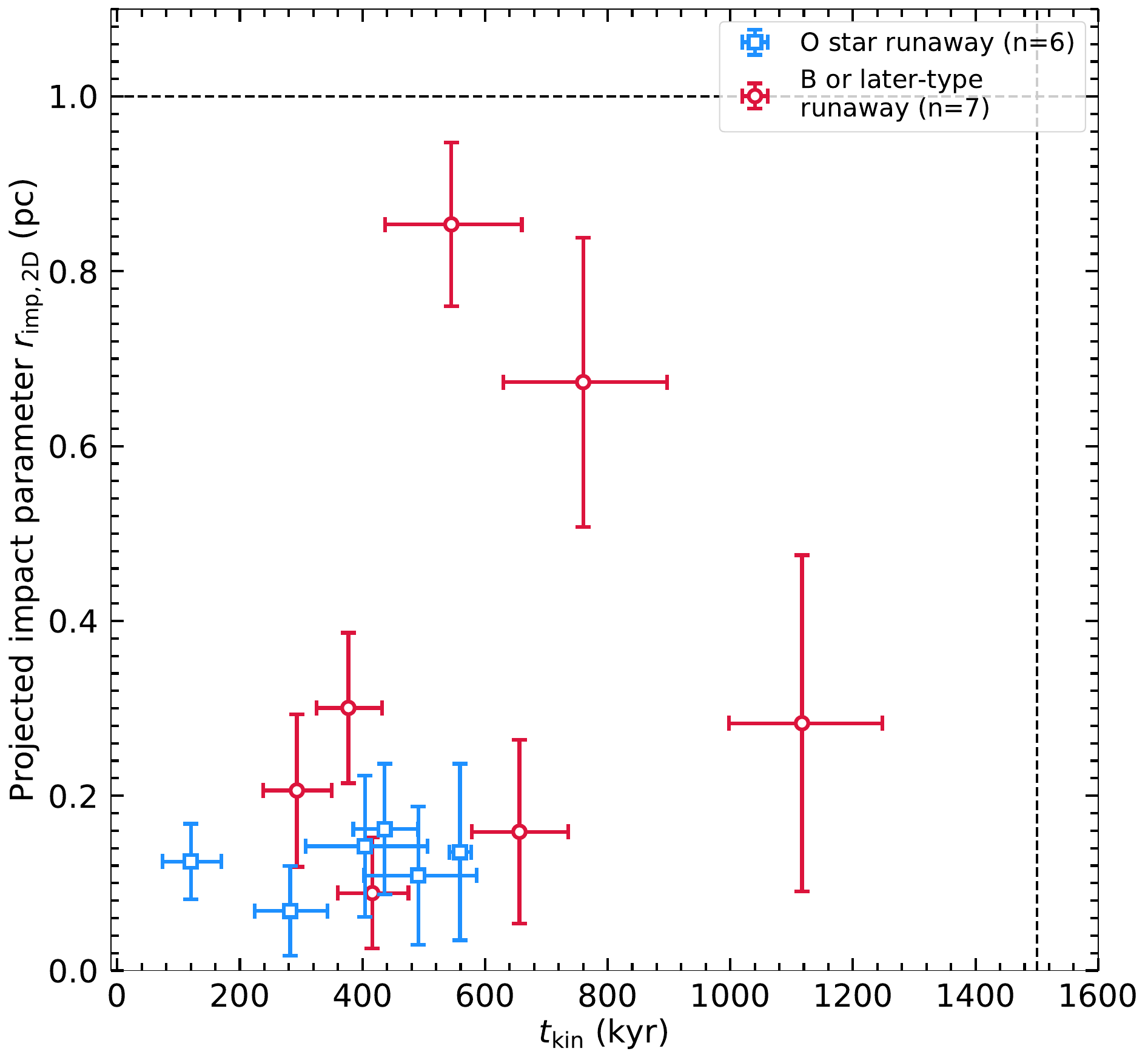}
\caption{Projected impact parameter ($r_{\rm{imp,2D}}$) of the runaways as a function of their kinematic age ($t_{\rm{kin}}$), marked and coloured similarly as in Figure~\ref{fig:tkin_runaways}. The cut-off $r_{\rm{imp,2D}}$ and $t_{\rm{kin}}$ used in the runaway search are indicated with the dashed black lines.}
\label{fig:rimp_tkin_runaways}
\end{figure}

Last, Figure~\ref{fig:rimp_tkin_runaways} shows $r_{\rm{imp,2D}}$ as a function of $t_{\rm{kin}}$. We have converted this quantity to the physical projected separation at the distance of NGC\,6618 assuming a negligible radial motion. Almost all runaways are consistent with being ejected within 0.2-0.4 pc of the centre of NGC\,6618. The identified O stars could all have been ejected from within $\sim$ 0.25 pc. Two outliers can be seen which are determined to have $r_{\rm{imp,2D}}$ between 0.5-1.0 pc. These two runaways are also ejected relatively long ago compared to the other runaways and are also relatively faint with $M_{\rm{K_{s}}}$ between --0.5 and --1.0. It could be possible that these were ejected from a different part of the cluster or represent false positives, and we have searched too far back in time and at too faint $M_{\rm{K_{s}}}$ magnitudes. Nevertheless, we have included them for completeness.


\begin{table*}
\centering
\caption{Properties of runaways coming from NGC\,6618, sorted by increasing kinematic age $t_{\rm{kin}}$.}
\label{tab:runaways}
\begin{tabular}{l l l l l l l l l}
\hline
\hline
\noalign{\smallskip}Identifier & K$_{\rm{s}}$ & Spectral type & |$\Delta v_{\rm{T}}$| & $v_{\rm{R}}$ & $t_{\rm{kin}}$ & $r_{\rm{imp,2D}}$ & A$_{\rm{V}}$ & Ref.\\
\noalign{\smallskip}- & mag & - & km s$^{-1}$ & km s$^{-1}$ & kyr & arcmin & mag & - \\
\noalign{\smallskip}\hline
\noalign{\smallskip}SLS373 & 6.8 & \textit{O3-6: V}\tablefootmark{a} & 20.2 $\pm$ 1.5 & - & 120$^{+50}_{-46}$ & 0.255 $\pm$ 0.091 & $\sim$ 9 & 1 \\
\noalign{\smallskip}SLS17 & 7.8 & \textit{O7-B0: V}\tablefootmark{a} & 18.8 $\pm$ 0.9 & - & 282$^{+61}_{-58}$ & 0.14 $\pm$ 0.10 & $\sim$ 7 & 1 \\
\noalign{\smallskip}OI 672 & 9.8 & \textit{B1-3: V}\tablefootmark{a} & 18.4 $\pm$ 0.9 & -29 $\pm$ 9\tablefootmark{b} & 293$^{+58}_{-55}$ & 0.42 $\pm$ 0.17 & $\sim$ 6 & 2 \\
\noalign{\smallskip}2MASS J18205435-1556291 & 10.6 & \textit{B5-9: V}\tablefootmark{a} & 19.9 $\pm$ 0.7 & - & 377$^{+55}_{-52}$ & 0.61 $\pm$ 0.18 & $\sim$ 4 & - \\
\noalign{\smallskip}LS 4943 & 8.9 & O9.7 V & 10.7 $\pm$ 0.4 & - & 404$^{+103}_{-97}$ & 0.30 $\pm$ 0.16 & $\sim$ 3 & 1 \\
\noalign{\smallskip}BD-164832 & 8.9 & B0 V + B1 V & 19.0 $\pm$ 0.7 & - & 416$^{+59}_{-56}$ & 0.18 $\pm$ 0.13 & $\sim$ 2 & - \\
\noalign{\smallskip}LS 4941 & 9.5 & O9.7 V & 20.4 $\pm$ 0.6 & - & 436$^{+54}_{-51}$ & 0.33 $\pm$ 0.15 & $\sim$ 3 & - \\
\noalign{\smallskip}BD-164826 & 7.3 & O5 V((f))z + O9/B0 V & 12.4 $\pm$ 0.5 & 11 $\pm$ 1 & 491$^{+95}_{-89}$ & 0.22 $\pm$ 0.16 & $\sim$ 4 & 3, 4 \\
\noalign{\smallskip}OI 637 & 10.3 & \textit{B2-5: V}\tablefootmark{a} & 10.0 $\pm$ 0.4 & -14 $\pm$ 6\tablefootmark{b} & 545$^{+127}_{-119}$ & 1.75 $\pm$ 0.19 & $\sim$ 3 & 2 \\
\noalign{\smallskip}2MASS J18182392-1721517\tablefootmark{c} & 7.9 & \textit{O7-B0: V}\tablefootmark{a} & 65.3 $\pm$ 2.8 & - & 559$^{+18}_{-17}$ & 0.28 $\pm$ 0.20 & $\sim$ 6 & - \\
\noalign{\smallskip}2MASS J18194850-1626446 & 10.6 & \textit{B2-6: V}\tablefootmark{a} & 13.2 $\pm$ 0.4 & - & 656$^{+81}_{-78}$ & 0.33 $\pm$ 0.21 & $\sim$ 4 & - \\
\noalign{\smallskip}LS 4951 & 10.1 & \textit{B1.5-4: V}\tablefootmark{a} & 10.9 $\pm$ 0.4 & - & 761$^{+148}_{-141}$ & 1.37 $\pm$ 0.33 & $\sim$ 4 & 1 \\
\noalign{\smallskip}TYC 6265-1828-1
 & 10.5 & \textit{B2-6: V}\tablefootmark{a} & 9.5 $\pm$ 0.4 & 9 $\pm$ 27\tablefootmark{b} & 1117$^{+124}_{-130}$ & 0.55 $\pm$ 0.36 & $\sim$ 2 & - \\
\noalign{\smallskip}\hline
\end{tabular}
\tablefoot{
\tablefoottext{a}{Spectral type estimated from photometry;}
\tablefoottext{b}{Radial velocity from \textit{Gaia} DR3;}
\tablefoottext{c}{Bowshock in \textit{Spitzer} MIPS}
}
\tablebib{(1) \citet{Povich2009}; (2) \citet{Ogura1976}; (3) \citet{MaizApellaniz2019}; (4) \citet{Williams2013}.}
\end{table*}

\begin{table}
\centering
\caption{Known and candidate OB stars in NGC\,6618\,PG and M\,17\,EB.}
\label{tab:NGC6618PG}
\begin{tabular}{l l l l l}
\hline
\hline
Identifier & Sp. type & K$_{\rm{s}}$ & Distance & Runaway \\
- & - & mag & pc & - \\
\hline
\multirow{2}{*}{BD--16 4826} & O5 V((f))z & \multirow{2}{*}{7.3} & \multirow{2}{*}{1741$^{+58}_{-48}$} & \multirow{2}{*}{T} \\ \vspace{1mm}
& + O9/B0 V & & & \\ \vspace{1mm}
BD--16 4831 & O9.7 Ia & 7.5 & 2437$^{+146}_{-107}$ & N \\ \vspace{1mm}
BD--16 4834\tablefootmark{a} & O9.5 II & 7.9 & - & N \\ \vspace{1mm}
BD--16 4822 & B2.5 II & 8.2 & 1593$^{+49}_{-41}$ & N \\ \vspace{1mm}
\multirow{2}{*}{BD--15 4928} & B0.5 V & \multirow{2}{*}{8.3} & \multirow{2}{*}{1676$^{+63}_{-52}$} & \multirow{2}{*}{F} \\ \vspace{1mm}
& + B1.5 V & & & \\ \vspace{1mm}
\multirow{2}{*}{LS 4972} & B1 V & \multirow{2}{*}{8.7} & \multirow{2}{*}{1591$^{+50}_{-42}$} & \multirow{2}{*}{F} \\ \vspace{1mm}
& + B2 V & & & \\ \vspace{1mm}
LS 4952 & \textit{O9-B2: V}\tablefootmark{b} & 8.8 & 1988$^{+92}_{-72}$ & N \\ \vspace{1mm}
LS 4943 & O9.7 V & 8.9 & 1591$^{+47}_{-40}$ & Y \\ \vspace{1mm}
\multirow{2}{*}{BD--16 4832} & B0 V & \multirow{2}{*}{8.9} & \multirow{2}{*}{1630$^{+55}_{-46}$} & \multirow{2}{*}{T} \\ \vspace{1mm}
& + B1 V & & & \\ \vspace{1mm}
P09 \#25\tablefootmark{a,c} & - & 9.4 & - & N \\ \vspace{1mm}
LS 4941 & O9.7 V & 9.5 & 1562$^{+41}_{-36}$ & Y \\ \vspace{1mm}
LS 4970 & \textit{B0-3: V}\tablefootmark{b} & 9.6 & 1959$^{+93}_{-72}$ & N \\ \vspace{1mm}
LS 4949 & \textit{B1-4: V}\tablefootmark{b} & 9.8 & 1835$^{+105}_{-78}$ & N \\ \vspace{1mm}
LS 4951 & \textit{B1-7: V}\tablefootmark{b} & 10.1 & 1601$^{+50}_{-42}$ & Y \\
\hline
\hline
\end{tabular}
\tablefoot{
\tablefoottext{a}{\texttt{ruwe} $>$ 1.4;}
\tablefoottext{b}{Spectral type estimated from photometry;}
\tablefoottext{c}{Also known as \textit{2MASS} J18204652-1537559.}
}
\end{table}

\section{Discussion}
\label{sec:discussion}

\subsection{Distance to M\,17}
The distance to M\,17 has been debated in the literature for decades, with estimates ranging from 1.3 to 2.9 kpc \citep{Ogura1976,Hanson1997,Russeil2003}. More recent distance estimates resulted in $\sim$ 1.6 and 2.1 kpc \citep[][]{Povich2007,Hoffmeister2008}. The distance of 1.98$^{+0.14}_{-0.12}$ kpc by \citet{Xu2011} to a maser associated with M\,17 lies in between the latter two distance estimates and settled these discrepancies. We note, however, that \citet{Xu2011} fit three astrometric parameters (parallax and proper motions) to only four data points, leaving room for observational biases.

With \textit{Gaia}, the astrometric distance to M\,17 is closer to $\sim$ 1.7 kpc, accounting for the parallax zero-point offset. \citet{Kuhn2019} estimates a distance of 1.68$^{+0.13}_{-0.11}$ kpc with DR2, \citet{MaizApellaniz2022a_villafranca} recover 1.70$^{+0.041}_{-0.039}$ kpc with EDR3 and we find 1.674$^{+0.019}_{-0.018}$ kpc with DR3. \citet{Kuhn2021} estimate a distance of 1.54$^{+0.16}_{-0.20}$ kpc, from a sample of 11 member stars. The distance determined by \citet{Xu2011} is consistent within 2$\sigma$ with a distance of 1.75 kpc, not including the aforementioned uncertainties.

The recent \textit{Gaia} based estimates for the distance of NGC\,6618 differ by a factor of $\sim$ 0.85 from the previously used maser estimate \citep[see e.g.][]{RamirezTannus2017}. This has implications for luminosity and mass estimates of young stellar objects in NGC\,6618. Masses estimated from spectral-energy-distribution fitting and spectroscopy scale with $d^{2}$, which results in mass estimates for these objects decreasing on average by a factor of $\sim$ 0.71.

\subsection{Runaway properties}
We have identified 13 runaway systems consistent with being ejected from the centre of NGC\,6618. BD--16 4826 is a known binary system of which the primary is an O5 V((f))z star and the secondary could be either an O9 or B0 V star. Six to seven of these systems are likely O-type stars, while six to seven others are likely to be B-type stars. We note several observational biases. We have introduced a brightness cut-off at $M_{\rm{K_{s}}}$ $<$ 0 mag, which is why no runaways were found with A or later spectral types. Similarly, we could miss out on heavily extincted, therefore fainter runaways either due to such systems having large uncertainties in the \textit{Gaia} astrometry or simply because they are not detected at all.

We can estimate the percentage of massive runaways in two ways. First, we compare the number of O-type runaways to the total number of O stars originating from NGC\,6618. We have six O-type runaways and 15 O-type stars located in NGC\,6618, which includes the runaways and excludes the background O9.7 Ia star. The O star runaway percentage is therefore 29\%. Not only could we miss O-type runaways due to the aforementioned biases, we could also miss O-type stars in NGC\,6618 because of the extreme extinction in the cluster itself. The 29\% of O-type runaways is therefore still somewhat uncertain.

Second, we use a brightness cut-off to estimate the percentage of massive runaway stars. The detectability of massive stars in NGC\,6618 diminishes for $\rm{K_{s}}$-magnitudes fainter than $\gtrsim$ 9.0 mag. We have seven runaways and 22 stars in NGC\,6618 brighter than this. This results in a similar runaway percentage of 32\% for stars brighter than $\rm{K_{s}}$ $\lesssim$ 9.0 mag. We have included two systems for which their origin is unclear (BD-16 4822 and LS 4972) as part of NGC\,6618.

Since these runaways originate from a cluster with an age estimated to be $\lesssim$ 1 Myr \citep{Hanson1997,Hoffmeister2008,Povich2009,RamirezTannus2017}, significantly less than the lifetime of a massive star, they should have been ejected by dynamical interactions. We compare the percentage of ejected O stars to other clusters where dynamical ejections have been observed and to the runaway percentage in numerical simulations of star clusters. Dynamically ejected O-type runaways have been observed for an increasing number of young massive clusters. In-depth investigations on these runaways with \textit{Gaia} have now been done for Westerlund 2, NGC 3603, NGC 7000, the Orion Nebula Cluster and NGC 6611 \citep{Drew2018,Drew2019,MaizApellaniz2022b_bermuda,Farias2020,Stoop2023}. Westerlund 2 is estimated to have eight O star runaways, and considering that there are still 30 O stars in the centre of the cluster, this would translate to an O star runaway percentage of $\sim$ 21\% \citep{Vargas2013,Drew2021}. On the one hand, undiscovered O + O binaries may cause this percentage to be overestimated. On the other hand, more O star runaways might exist with velocities below the adopted threshold of $|\Delta v_{\rm{T}}|$ $\gtrsim$ 20 km s$^{-1}$. Our adopted runaway velocity threshold of $>$ 5 km s$^{-1}$ makes direct comparisons more difficult. We have discovered four O star runaways with $|\Delta v_{\rm{T}}|$ $\gtrsim$ 20 km s$^{-1}$, yielding a runaway percentage of $\sim$ 17\% above this velocity.

NGC\,6618 is similar in O star content to NGC 6611 ($\sim$ 21 compared to 19 respectively). Adopting the same $|\Delta v_{\rm{T}}|$ threshold of $\gtrsim$ 5 km s$^{-1}$, we obtain 4-5 O star runaways in NGC 6611 (or $\sim$ 21-26\%). Several O stars in NGC 6611 are more likely to be radial runaways \citep[see table 2 in][]{Stoop2023}, and \citet{MaizApellaniz2018} also note that up to 50\% of the runaways may not be found if only considering the proper motions. The analysis of Westerlund 2, NGC 6611 and NGC\,6618 suggest that O star runaway percentages of 15-20\% are typical for 2D velocities above 20 km s$^{-1}$, but that a considerable fraction of O stars can also have 2D velocities of 3-20 km s$^{-1}$. In the case of NGC 6611 and NGC\,6618, another 15-20\% may have 2D velocities in this range. In order to make more robust statements on the fraction of runaways and their velocities, more detailed investigations are needed.

Almost all runaways move with transverse velocities of $\sim$ 10 to 20 km s$^{-1}$, except for the high-velocity runaway 2MASS J18182392-1721517 with $|\Delta v_{\rm{T}}|$ $\sim$ 65 km s$^{-1}$. It is unclear why these 12 runaways have such similar transverse velocities. There could be a common physical origin where the initial star formation, stellar density and stellar dynamics resulted in similar ejection velocities. This is speculative at best and we note that the radial motion of these runaways have not been taken into account yet and that the true velocity distribution could look different.

\subsection{Constraints for the dynamical calculations of young massive clusters}
The runaway properties can be compared to dynamical simulations of young massive clusters to constrain cluster properties such as the initial radius, mass segregation, binary fraction, separation, and mass ratio \citep{Fujii2011,Oh2016}. We compare our results to the dynamical simulations of \citet{Oh2016}. The ejection fraction of O stars ($\sim$ 0.3) agrees best with models that assume a relatively small half-mass radius $r_{\rm{h}}$(0) of $\sim$ 0.3\,pc and a high initial binary fraction\footnote{Our results agree best with the five models \texttt{MS3OP\_SPC}, \texttt{MS3OP\_SP}, \texttt{MS3UQ\_SP}, \texttt{MS3OP}, and \texttt{NMS3OP} in \citet{Oh2016}}. Moreover, most of these models assume an initial mass segregation, with more massive stars located closer to the cluster centre, and mass-ratios (between the secondary and primary star) that tend to unity. The best value $r_{\rm{h}}$(0) agrees well with our results that the O stars were all ejected from within the inner $\sim$0.1--0.2\,pc. The one model with $r_{\rm{h}}$(0) = 0.1\,pc has a significantly higher O star ejection fraction of $\sim$ 0.52. Models with $r_{\rm{h}}$(0) = 0.8\,pc can not produce O star ejection fractions comparable to our results. A pairing of primary and secondary masses that is random causes the massive stars to preferentially be in binaries with lower mass stars, as these dominate the total number of stars. Such a pairing is unfavourable relative to an ordered pairing, which produces mass ratios that are close to unity.

The peak of the velocity distribution of the ejected O stars in the best fitting \citet{Oh2016} models are in excellent agreement with most of the observed transverse velocities of $\sim$10 to 20 km s$^{-1}$. These models predict that most O stars are ejected 0--1 Myr after the start of the simulation, but some may be ejected at late as 2--3 Myr. Interestingly, \citet{Oh2016} find that the binary properties of the ejected O stars differ for these five models. Their multiplicity fraction ranges from $\sim$ 0.1--0.2 for model \texttt{MS3OP} to $\sim$ 0.4--0.6 for \texttt{MS3OP\_SP}. The orbital period, mass ratio, and eccentricity distribution of the runaway population also differ significantly for the five models. \texttt{MS3OP\_SP} and \texttt{MS3OP\_SPC} have typical mass ratios between 0.8 and 1.0 as a result of the ordered pairing of massive binaries. We have observed two runaway binaries, BD--16 4832 and BD--16 4826, the latter has an orbital period $\Porb \sim 16$ days. The mass ratio of these two binaries are $\sim$ 0.6--0.8 and 0.5, respectively. Due to the small number statistics, we can not draw definitive conclusions on which specific model setup yields the best fit. At face value, our results best match model \texttt{MS3UQ\_SP}. This model has $r_{\rm{h}}$(0) = 0.3\,pc, assumes primordial mass segregation for massive stars, a uniform binary mass fraction distribution, and the \citet{Sana2012} orbital period distribution for massive stars.

More investigation is needed regarding the dynamical evolution of young massive clusters to draw conclusive statements on their initial properties. The models of \citet{Oh2016} are tailored for young massive clusters with $\sim$ 10 O stars initially, while NGC\,6618 formed $\sim$ 21 O stars. The binary properties of both the O stars currently in the cluster and runaways could be key in uncovering the initial conditions of massive binaries. The binary properties such as the \Porb, mass ratio, and eccentricity help break degeneracies in dynamical simulations. \citet{RamirezTannus2021} find that massive binaries have significantly wider orbits in the 1-2 Myr after birth compared to the massive binaries typically studied in massive clusters with ages = 2-6 Myr \citet{Sana2012}. This implies that the \citet{Sana2012} \Porb\ distribution adopted in N-body simulations such as in \citet{Oh2016} may need to be adjusted to an initially wider configuration. This could alter the O star runaway fraction, velocity distribution, and binary runaway properties.

We list here our findings which may help detailed numerical simulations of star clusters to reproduce our findings. There is a period of $\sim$ 500 kyr during or shortly after star formation where 30\% of the O stars are ejected. The binary fraction of these runaways is not 0, as two of the four spectroscopically observed runaways are found to be binaries. The O star runaways should all be ejected within 0.2-0.3 pc from the centre of the cluster. The 2D velocity of most the runaways is found to be in the range between 10-20 km s$^{-1}$, and a fast runaway with a 2D velocity $>$ 30 km s$^{-1}$ is also found. The masses of the runaways are higher than $\gtrsim$ 3 M$_{\odot}$ considering our cut-off at K$_{\rm{s}}$ $<$ 0 mag \citep{Pecaut2013}.

\subsection{Age of NGC\,6618}
NGC\,6618 is especially interesting because it is suggested to be one of the youngest clusters known in our Galaxy. We have identified runaways with a kinematic age ranging from $\sim$ 100 to 1250 kyr. The O star runaways specifically have a kinematic age between $\sim$ 100 and 600 kyr. Simulations of young star clusters show that dynamical ejections commence during and right after star formation \citep{Bate2002,Fujii2011,Oh2016}. The kinematic ages of these runaways therefore convey information about when NGC\,6618 was at its densest. The runaways that were ejected first may have been lost during or closely following the star formation process. We propose that we can use the kinematic age of the first runaways to estimate the age of NGC\,6618. To refrain from using a single measurement to determine the age, we split the runaway sample into two halves based on their kinematic ages. The second sample contains the seven runaways with the largest kinematic ages, which range from $\sim$ 400 to 1250 kyr. We perform a Monte Carlo simulation to take the uncertainty on the kinematic age into account. For each runaway, a random $t_{\rm{kin}}$ is drawn between its lower and upper bound assuming a uniform distribution. We calculate the mean and standard deviation on these $t_{\rm{kin}}$ in each iteration. The age and uncertainty are given by the 50$^{\rm{th}}$ percentile of the means and the 84$^{\rm{th}}$ percentile of the standard deviations, respectively. This results in an age of NGC\,6618 equal to 0.65 $\pm$ 0.25 Myr.

This method of using the kinematic age of the runaways to match to the age of a cluster also works well for several other young clusters. \citet{Stoop2023} show that for NGC 6611 the isochrone age = 1.3 $\pm$ 0.2 Myr. Using the 50\% of runaways with the largest kinematic ages in their table 2 and the method described above, yields an age of 1.35 $\pm$ 0.31. For the Orion Nebula Cluster (ONC), the age of the oldest population is within 1$\sigma$ uncertainties 2.51-3.28 Myr \citep{Beccari2017}. The runaways $\mu$ Colombae and AE Aurigae were ejected $\sim$ 2.5 Myr ago from the ONC, consistent with the age of their natal cluster. \citet{Kroupa2018} also propose that $\mu$ Colombae and AE Aurigae were ejected during or shortly after star formation to explain the two younger populations found in the ONC. NGC 7000 also ejected several massive stars 1.5-2.0 Myr ago, consistent with the isochronal age of $\sim$ 1.8 Myr \citep{Kuhn2020, MaizApellaniz2022b_bermuda}.

The runaways in NGC\,6618 could only have been ejected if the `bully' stars were already present \citep{Fujii2011}. The kinematic age of the first ejected runaway also gives a lower limit on when the first stars formed in the cluster. The first ejected O star is the high velocity runaway \textit{2MASS} J18182392-1721517 with $t_{\rm{kin}}$ between 0.54-0.58 Myr. This suggests that the first stars in NGC\,6618 formed at least longer than 0.54 Myr ago, which is in agreement with the previously determined age = 0.65 $\pm$ 0.25 Myr. While there are three B-type runaways with a larger kinematic age, it is unclear if they truly came from NGC\,6618. They have lower transverse velocities, making it harder to determine their kinematic ages. Two of these were also ejected further away from the centre of NGC\,6618 (0.5-1.0 pc), making their origin less clear. They could have been ejected from a neighbouring star-forming sub-cluster, or they could be interloper field stars. O stars are relatively rare, and almost all originate from young massive clusters or OB associations \citep{Gies1987,deWit2005}. We can therefore be more confident that \textit{2MASS} J18182392-1721517 is a true runaway coming from NGC\,6618.

\subsection{Physical properties of NGC\,6618}
NGC\,6618 is reported to contain at least 15 O stars \citep{Povich2009,RamirezTannus2017}. If we include the O star runaways and several O stars `hidden' in the nearby field, we instead have 21 O stars total born in NGC\,6618. We could still miss out on more O stars shrouded by extinction, or hidden in binaries. The O star population of NGC\,6618 may therefore be at least 30 to 40\% larger than initially thought. We can make a rough estimate of the cluster mass (M$_{\rm{cl}}$) by extrapolating an initial mass function (IMF) down to lower masses. We adopt here the \citet{Kroupa2001} IMF and assume initially 21 O stars with masses $\gtrsim$ 18 M$_{\odot}$. Integrating the IMF down to the brown dwarf lower limit of 0.08 M$_{\odot}$, we obtain M$_{\rm{cl}}$ $\sim$ 5.1 $\times$ 10$^{3}$ M$_{\odot}$. We note that there is a significant uncertainty in this calculation since we are extrapolating the small number of O stars down to the lower mass stars, which make up most of the mass of a cluster.

The runaway O stars are typically not accounted for in the mass function of young massive clusters. It may be that the these resemble a Kroupa mass function before accounting for runaway O stars. The fraction of O star runaways is thought to be significantly higher than for B or later-type stars \citep{Gies1987}. If these runaway O stars are accounted for in the mass function, this may result in a significantly more top-heavy distribution than commonly assumed. This has also been noted by \citet{MaizApellaniz2022b_bermuda}, who find that NGC 7000 may also have a top-heavy IMF if the runaway O stars are accounted for. \citet{Schneider2018} also find a top-heavy IMF for the 30 Doradus region in the Large Magellanic Cloud, where the power-law exponent of the IMF was determined to be $\sim$ 1.9 for masses above 15 M$_{\odot}$. Since this study focused on the entire surrounding region and not on the R136 star cluster, there were almost no biases regarding either the inclusion or exclusion of runaways. Larger sample studies on a top-heavy Galactic IMF are needed to investigate whether indeed the IMF features a kink to a shallower slope above $\sim$ 15 M$_{\odot}$.

\subsection{Previous star formation}
\label{sec:disc_ngc6618pg}

\begin{figure*}
\centering
\includegraphics[width=0.99\linewidth]{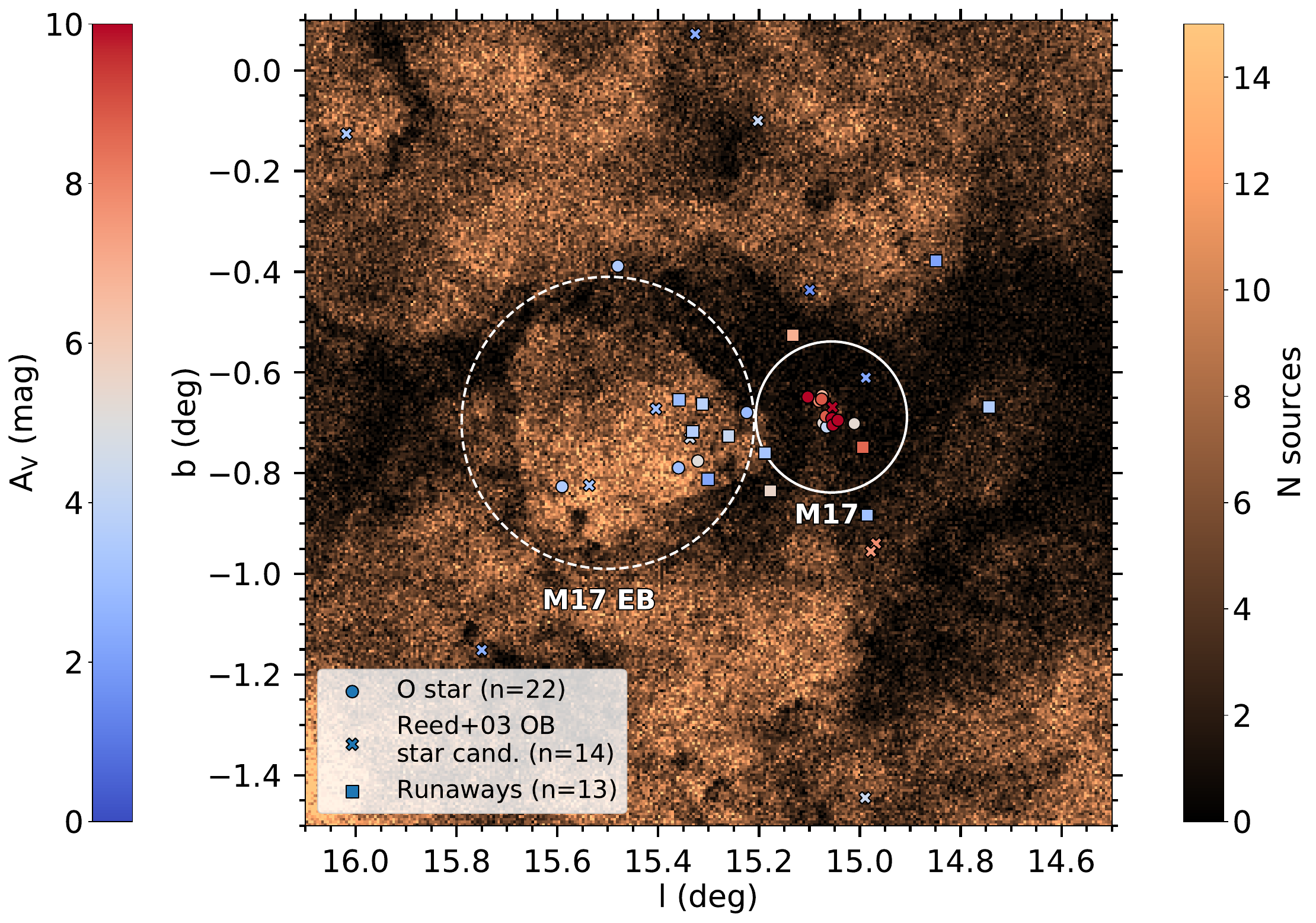}
\caption{Large-scale spatial distribution of the O stars and runaways originating from NGC\,6618. We include for context the candidate OB stars in the \citet{Reed2003} catalogue. Their markers are coloured according to their A$_{\rm{V}}$ with the colour-bar on the left. For the entire field, we show a number-density map coloured according to the \textit{Gaia} source density with the colour-bar on the right. Each pixel is 1.2$^{\arcmin}$ by 1.2$^{\arcmin}$ in size. We show for context the location of M\,17 and M\,17\,EB with the solid and dashed white circle, respectively.}
\label{fig:lb_density}
\end{figure*}
The presence of several massive stars, excess soft X-ray emission and an extended bubble have been used as evidence for a generation of older stars making up the progenitor OB association NGC\,6618\,PG \citep{Povich2007,Povich2009}. To further investigate this, we have collected all stars which could be part of NGC\,6618\,PG in Table~\ref{tab:NGC6618PG}. These are made up of the stars listed in \citet{Povich2009} and possible OB stars in \citet{Reed2003} located inside the M\,17\,EB region. We have excluded stars that we now know to have a late spectral type. We list their K$_{\rm{s}}$ magnitude, distance of the selected stars, and whether we have identified these stars as a runaway from NGC\,6618 instead.

We have shown that four of the seven massive stars that could be part of NGC\,6618\,PG are instead runaways coming from NGC\,6618. One of the dominant ionising sources, the now spectrally classified O9.7 Ia star BD--16 4831, is likely not associated with either NGC\,6618 or NGC\,6618\,PG as this star is located significantly further away at a distance of $\sim$ 2.4 kpc. The remaining O star BD--16 4834 has poor astrometry and we can not draw any conclusions to which cluster this star might belong. The presence and extent of NGC\,6618\,PG is thus significantly less clear than initially thought. NGC\,6618\,PG would then consist of only B or later-type stars, which will have significant implications for its age. If the O stars do not make up NGC\,6618\,PG, it becomes less clear what caused M\,17\,EB.

Instead, we investigate whether the O star runaways ejected in the direction of the M\,17\,EB could have caused this faint diffuse \HII region \citep{Povich2009}. We can assume that the ionising power of all runaways is dominated by the O5 V star in BD--16 4826, as the most massive star easily dominates the ionising budget. This gives log($Q_{0}$) = 49.26 ionising photons per second \citep{Martins2005}. The ionising power of the other runaways combined, which are all O9-B1 II/V, is still an order of magnitude less than the O5 V star. We assume a Str\"{o}mgren sphere with a radius $R_{\rm{S}}$ $\sim$ 7.5 pc similar to \citet{Povich2009} and a recombination coefficient $\alpha_{B}$ of 2.57 $\times$ $10^{-13}$ cm$^{-3}$ s$^{-1}$ \citep{Storey1995}. We can calculate what the particle density $n$ in the Str\"{o}mgren sphere must be, which is given by
\begin{align}
    n_{\rm{ISM}} = \sqrt{\frac{3 Q_{0}}{4 \pi \alpha_{A} R^{3}_{S}}}.
\end{align}
Given these values, we determine $n_{\rm{ISM}}$ of the order of 40 cm$^{-3}$. \citet{Povich2009} determine $n_{0}$ $\sim$ 350 cm$^{-3}$ from the dynamics of the CO and molecular gas, which includes two denser surrounding molecular clouds: M\,17 North and part of MC G15.9-0.7. Our estimate of $n$ lies in between typical particle densities of the interstellar medium ($\sim$ 1 cm$^{-3}$) and the $n_{0}$ determined by \citet{Povich2009}. Our estimated $n_{\rm{ISM}}$ is therefore not surprising, given that the \HII region lies in between the two aforementioned molecular clouds.

We show the spatial distribution of the OB stars on a larger scale in Figure~\ref{fig:lb_density} in the direction of M\,17\,EB. We have included the candidate OB stars catalogued in \citet{Reed2003} in the field around M\,17\,EB as well for context. The only constraint we have put on the \citet{Reed2003} candidate OB stars is that they should be located between 1.5 and 2.0 kpc to exclude obvious fore and background stars.

Still, an over-density of OB stars is present at the location of NGC\,6618\,PG, even if we account for five runaways. We also show in Figure~\ref{fig:lb_density} for the entire field a 2D histogram of all \textit{Gaia} sources, coloured according to their source density and a colour-bar on the right. Each pixel is 1.2$^{\prime}$ by 1.2$^{\prime}$, with brighter colours indicating more sources than darker colours. If we compare Figure~\ref{fig:lb_density} to the extinction map in figure 9 in \citet{Povich2009}, we see that these \textit{Gaia} source densities match the extinction map well. The \textit{Gaia} source density can be used as an analogue to an extinction map, where brighter colours indicate less extinction so that more sources are visible, and vice versa for darker colours.

The dark lanes around M\,17 highlight the extreme extinction found in this region ($A_{\rm{V}}$ $>$ 5 mag), so that few sources are visible with \textit{Gaia} in the optical. The M\,17\,EB shows significantly more sources than the dark lanes surrounding it. If the ionising star in BD--16 4826 has carved out this \HII region, it could have decreased the extinction in this line of sight compared to the neutral medium surrounding M\,17\,EB. 

We have estimated the A$_{\rm{V}}$ for all runaways, OB star candidates in \citet{Reed2003}, and O stars in NGC\,6618\,(PG) from their \textit{2MASS} J -- K colour, for which the intrinsic colour is $\sim$ 0.24 mag for OB stars assuming the \citet{Cardelli1989} extinction law with R$_{\rm{V}}$ = 3.1. The markers of these stars are coloured according to their A$_{\rm{V}}$ in Figure~\ref{fig:lb_density}, with bluer colours depicting less extinction compared to redder colours with more extinction. The O stars in NGC\,6618 can be seen to be reddened (A$_{\rm{V}}$ $>$ 5 mag), as is observed \citep[e.g.][]{RamirezTannus2017}. 

The sources in NGC\,6618\,(PG) suffer from less extinction, depicted by their bluer colours (A$_{\rm{V}}$ $<$ 5 mag). For example, we estimate that the most blue system BD--16 4832 (B0 V + B1 V) in Figure~\ref{fig:camd}, located in M\,17\,EB only has $A_{\rm{V}}$ $\sim$ 2 mag. With less extinction in this line of sight, the early to mid-type B stars are more easily visible and picked up by OB star catalogues such as in \citet{Reed2003}. Conversely, fewer OB star candidates are detected by both \textit{Gaia} and other catalogues in the dark lanes surrounding NGC\,6618, which is why we may miss runaways and OB stars in these directions. This can explain why we observe more runaways in the direction of M\,17\,EB., since we expect runaways to be ejected isotropically. The total number of runaways with stellar masses $>$ 3 M$_{\odot}$ could therefore be intrinsically larger and our observed 13 runaways may be a lower limit. We observe six to seven runaways in the direction of M\,17\,EB. Only an additional four to five runaways were ejected in this quadrant (east) compared to other quadrants, which could also be explained by low number statistics. This does not imply that NGC\,6618\,PG does not exist and that all the O and B stars make up an `imposter' cluster. We list this as a possibility and note that NGC\,6618\,PG is likely less massive than initially thought, but may still exist.

\subsection{Bowshock}
\label{sec:disc_bowshock}
Not all runaways apparently produce a wind bow shock, which is also the case here \citep{Huthoff2002}. The runaway star 2MASS J18182392-1721517 can be seen to produce a bowshock in Figure~\ref{fig:bowshock}. We investigate whether 2MASS J18182392-1721517 is consistent with being a late-type O star. \citet{Baranov1971} show that momentum balances between the ram pressure exerted by the stellar wind and the ISM. For a given stellar wind velocity $v_{\rm{w}}$, a particle density of the ISM $n_{\rm{ISM}}$, a velocity of the star relative to the ISM $v_{*}$ and the distance between the star and the apex of the bowshock $R_{0}$, the stellar wind mass-loss rate should be
\begin{align}
    \dot{M}_{*} = \frac{4 \pi \ m_{\rm{H}}\ n_{\rm{ISM}}\ v^{2}_{*}\ R^{2}_{0}}{v_{\rm{wind}}}.
\end{align}
We assume $n_{\rm{ISM}}$ $\sim$ 1 cm$^{-3}$, typical for what is observed in the ISM. While we have determined $n_{\rm{ISM}}$ $\sim$ 30 cm$^{-3}$ in Section~\ref{sec:disc_ngc6618pg}, 2MASS J18182392-1721517 is located in an entirely different region than M\,17\,EB. Particle densities derived from bowshocks in our Galaxy are typically in the range of 0.1-10 cm$^{-3}$ and our assumed $n_{\rm{ISM}}$ seems a reasonable mean \citep{Peri2012,Peri2015}. We will take $v_{*}$ to be equal to the transverse velocity $|\Delta v_{\rm{T}}|$ = 65 km s$^{-1}$ and determine $R_{0}$ to be $\sim$ 30-60$^{\prime \prime}$ which translates to 0.25-0.5 pc at the distance of 2MASS J18182392-1721517 ($\sim$ 1.75 kpc). Finally, we assume $v_{\rm{wind}}$ $\sim$ 2000 km s$^{-1}$, which is typical for a late-type O star \citep{Prinja1990}.

We then estimate the mass-loss rate of 2MASS J18182392-1721517 and obtain $\dot{M}_{*}$ $\sim$ 1 $\times$ 10$^{-7}$ M$_{\odot}$ yr$^{-1}$, or log($\dot{M}_{*}$) = --7.0. This could easily vary by an order of magnitude given the uncertainties and assumptions. Nevertheless, this $\dot{M}_{*}$ is typical of mid-to late-type O stars \citep[see e.g.][]{Vink2000,Bjorklund2022}. The tentative spectral classification based on its distance and K$_{\rm{s}}$-magnitude is consistent with that needed to produce the bowshock. Optical spectra of this star could conform our findings.

\section{Summary \& conclusion}
\label{sec:conclusion}
We have studied NGC\,6618, located in M\,17, and it is one of the youngest massive clusters known in our Galaxy with \textit{Gaia} DR3. We summarise our findings here:
\begin{itemize}
    \item We have spectroscopically classified six new O stars, of which five are likely associated with NGC\,6618, in addition to the 17 known O stars in the cluster. In addition to this we have classified four B-type stars, of which two have been identified as SB2 B-type binaries.
    \item We have identified 42 members of NGC\,6618. M\,17 and NGC\,6618 are significantly closer than initially thought at a distance of 1.65-1.70 kpc. This has implications for studies investigating the rich pre-main-sequence population that is present.
    \item We have found 13 runaways coming from NGC\,6618, consistent with the dynamical ejection scenario. Out of the 13 runaways, six to seven are likely to be O stars and the remaining are likely to be B stars. The percentage of O stars ejected from NGC\,6618 is 27 to 35\%. These runaways are typically not accounted for in studies of the IMF of young clusters, which raises questions on whether the IMF could be more top-heavy than initially thought.
    \item The accurate \textit{Gaia} astrometry has allowed us to determine the transverse velocity, kinematic age, and impact parameter with great precision. The runaways have transverse velocities ranging from $\sim$ 10 up to 65 km s$^{-1}$. The kinematic ages range from $\sim$ 0.1 to 1.2 Myr, and all runaway O stars have kinematic ages between 0.1 to 0.6 Myr ago. Most of the runaways are ejected from inside 0.2 to 0.3 pc from the centre of NGC\,6618. 
    \item We find that the kinematic ages of the runaways can be used to estimate the ages of young clusters. Using the 50\% of runaways with the largest kinematic ages (ejected longest ago), we obtain an (kinematic) age = 0.65 $\pm$ 0.25 Myr. 
    \item We have discovered a bowshock in the \textit{Spitzer} MIPS 24$\mu$m image around one of the runaways 2MASS J18182392-1721517, which likely is a late-type O star.
    \item We find that the progenitor cluster north-east of NGC\,6618, NGC\,6618\,PG is likely to be less massive than initially thought. The main ionising sources in NGC\,6618\,PG are either runaways coming from NGC\,6618 or background stars. Instead, the emission bubble present at the location of NGC\,6618\,PG could have been caused by the ejected O stars.
\end{itemize}
M\,17 provides a unique opportunity to study a population of massive stars that has just arrived on the main sequence, to constrain the end of star formation properties, and to constrain the early history of dynamical interactions. Increasingly more accurate \textit{Gaia} astrometry will reveal more details about the young population present, including multiplicity properties of both the runaway and cluster stars.

\begin{acknowledgements}
The data and data products presented in this paper are available at the following DOI: \url{https://doi.org/10.5281/zenodo.8120575}. MS acknowledges support from NOVA. Some of the observations reported in this paper were obtained with the Southern African Large Telescope (SALT) under programme 2022-1-SCI-002 (PI: Paul Groot). Based on observations collected at the European Southern Observatory under ESO programme 0103.D-0099. This work has made use of data from the European Space Agency (ESA) mission
{\it Gaia} (\url{https://www.cosmos.esa.int/gaia}), processed by the {\it Gaia}
Data Processing and Analysis Consortium (DPAC,
\url{https://www.cosmos.esa.int/web/gaia/dpac/consortium}). Funding for the DPAC
has been provided by national institutions, in particular the institutions
participating in the {\it Gaia} Multilateral Agreement. This publication makes use of data products from the Two Micron All Sky Survey, which is a joint project of the University of Massachusetts and the Infrared Processing and Analysis Center/California Institute of Technology, funded by the National Aeronautics and Space Administration and the National Science Foundation.
\end{acknowledgements}

\clearpage

\bibliographystyle{aa}
\bibliography{Published.bib}

\clearpage

\begin{appendix}
\section{Spectroscopy, data reduction, and spectral classification}
\label{sec:appendix_spectroscopy}
We have obtained spectra for 10 systems in the O star sample in  either with the High Resolution Spectrograph (HRS) mounted on the Southern African Large Telescope (SALT) in Sutherland, the Intermediate-dispersion Spectrograph and Imaging System (ISIS) mounted on the William Herschel Telescope (WHT) at La Palma, or the medium resolution spectrograph X-shooter on the Very Large Telescope (VLT) at Paranal. We describe these observations and classify these stars based on the spectra below. The O stars spectrally classified here are listed first in Table~\ref{tab:member_Ostars}. Stars spectrally classified here as early B-type are stars at the bottom of the table.

\subsection*{SALT/HRS}
High-resolution spectroscopic observations have been taken of two stars, BD-16 4832 and BD-16 4834, with the HRS, a dual-beam fibre-fed, white-pupil, échelle spectrograph, in low resolution mode. The wavelength range is from 370-890\,nm, divided in a Blue and Red arm, at a resolving power of $R$\,$\sim$\,14,000. The spectra were taken as part of a monitoring campaign of three MYSOs in M\,17 (programme 2022-1-SCI-002). The data is reduced with the HRS MIDAS pipeline developed by \cite{Kniazev2016,Kniazev2017} and includes a flat field reduction, wavelength calibration and object subtraction. 

\subsection*{WHT/ISIS}
Eight stars have been observed with the ISIS optical spectrograph that includes a blue arm (388-478\,nm) and red arm (815-904\,nm). The slit width used for the observations is 0.79\arcsec\ resulting in a resolution of $R$\,$\sim$\,6000 for the blue arm and $R$\,$\sim$\,13,000 for the red arm. An ISIS reduction pipeline was developed in \textsc{python}, and allows for bias subtraction, flat fielding, overscan and cosmic ray removal, spectrum, and sky extraction and wavelength calibration, based on scripts from \cite{matt_craig_2017_1069648} and \cite{steve_crawford_2021_4588036}. 

\subsection*{VLT/X-shooter}
B181 has been observed with X-shooter, VLT, a multi-wavelength (300-2480\,nm) spectrograph with three separate spectroscopic arms. The resolving power of the arms is $R$\,$\sim$\,5400 in UVB arm (slit width 1.0\arcsec), $R$\,$\sim$\,8900 in the VIS arm (slit width 0.9\arcsec) and $R$\,$\sim$\,8100 in the NIR arm (slit width 0.6\arcsec). The data are taken as part of a multiplicity study in M\,17 (programme 0103.D-0099). The pipeline for X-shooter (esorex X-shooter 3.3.5) provided by ESO \citep{Modigliani2010} is used for flat field correction, bias subtraction and wavelength calibration. 

\subsection*{BD-16 4832}
The HRS spectrum of BD-164832 reveals a SB2 nature of the source. One of the two stars in the spectrum shows weak He\,{\sc ii} 4686\,\AA. Both stars have a strong He\,{\sc i} 4471\,\AA\ absorption compared to the Mg\,{\sc ii} 4481\,\AA, but for the star with the weak He\,{\sc ii} line, this ratio is slightly larger. Therefore the stars are classified as B0 + B1. The spectrum does not reveal signs of a luminosity class brighter than V (because of only weak O\,{\sc ii} 4350\,\AA\ absorption, close to H$\gamma$), however, close-by absorption features caused by an extended atmosphere are easier obscured in a SB2 spectrum.

\subsection*{BD-16 4834}
BD-164834 has an HRS spectrum that shows He\,{\sc i} and He\,{\sc ii} lines and where the relative strength between the He\,{\sc i} 4471\,\AA\ and He\,{\sc ii} 4542\,\AA\ points to a late O-type star. Since He\,{\sc i} 4144\,\AA\ is stronger than He\,{\sc ii} 4200\,\AA\ and the latter is slightly stronger than C\,{\sc iii} 4187\,\AA, the spectral type is about O9. This is narrowed down to O9.5 by the similar strength of He\,{\sc ii} 4686\,\AA\ and C\,{\sc iii} 4647/50/51\,\AA. The luminosity class is I-III by the strength of the N\,{\sc ii} 4379\,\AA\ compared to He\,{\sc i} 4387\,\AA. This is narrowed down to II by the relative strength of Si\,{\sc ii} 4116\,\AA\ to He\,{\sc i} 4121\,\AA\ and the lack of emission features. Therefore, BD-164834 is of spectral type O9.5 II.

\subsection*{BD-16 4831}
The ISIS spectrum of BD-164831 show weak He\,{\sc ii} lines pointing to a late O or early B spectral type. The classification based on photometry of \cite{Povich2009} is O4\,V or O7\,III. Since O\,{\sc ii} 4349\,\AA\ is not as strong as is expected for B-type stars, this is a late O-type star, more specifically O9.7 due to the similar strength of He\,{\sc ii} 4200\,\AA\ and C\,{\sc iii} 4187\,\AA. Based on the intensity of the Si\,{\sc iv} 4089\,\AA\ and Si\,{\sc iv} 4116\,\AA\ lines compared to hydrogen the star classifies as a supergiant. Since the emission lines of N\,{\sc iii} 4634/42\,\AA\ are absent, this is a luminous supergiant; therefore its classification is O9.7 Ia. 

\subsection*{BD-15 4928A}
This spectrum, taken with ISIS, shows strong He\,{\sc i} lines and one He\,{\sc ii} line at 4686\,\AA. This points to an early B type classification, different than the O8\,V classification from \cite{Povich2009}. He\,{\sc i} 4713\,\AA\ is stronger than He\,{\sc ii} 4686\,\AA\ and there is no Ni\,{\sc ii} 3995\,\AA\ which corresponds to a spectral classification of B0.5. The luminosity class is dwarf since O\,{\sc i} 4070/76\,\AA\ and Si\,{\sc ii} 4552/68/75\,\AA\ are not particularly strong. The classification of this star is B0.5\,V. 

\subsection*{BD-15 4928B}
BD-154928B is nearby BD-154928A on the sky as reported by \cite{Povich2009} with a spectral type B0.5\,V. This star has strong He\,{\sc i} lines but no He\,{\sc ii}, indicating an early B spectral type, but not B0. Since He\,{\sc i} 4144\,\AA\ is slightly stronger than He\,{\sc i} 4121\,\AA, the star is around spectral type B1. This is B1.5 because of a stronger He\,{\sc i} 4713\,\AA\ compared to the blended O\,{\sc ii} + C\,{\sc iii} 4650\,\AA\, which also points to a dwarf luminosity class. The spectral type of BD-154928B is B1.5\,V. 

\subsection*{B260}
B260 has an ISIS spectrum and is previously classified as O6-8 V \citep{Hanson1997, Hoffmeister2008, Povich2009}. The spectrum reveals strong He\,{\sc i} and He\,{\sc ii} lines pointing to a late O or B0-type star. The relative stronger He\,{\sc ii} 4686\,\AA\ compared to He\,{\sc i} 4713\,\AA\ suggests late O-type and the stronger He\,{\sc ii} 4542\,\AA\ compared to Si\,{\sc iii} 4553\,\AA, the latter being weaker than He\,{\sc i} 4387\,\AA\ suggests O9.5. The luminosity class is V based on the strength of C\,{\sc iii} 4647/50/51\,\AA, and the strength of the Si\,{\sc ii} 4116\,\AA\ line compared to He\,{\sc i} 4121\,\AA. The classification of this star is O9.5 V. 

\subsection*{LS4941}
The ISIS spectrum of LS4941 shows He\,{\sc i} and He\,{\sc ii} lines and a strong He\,{\sc ii} 4686\,\AA\ compared to He\,{\sc i} 4713\,\AA, ruling out the classification of \cite{Povich2009} of B1 V and instead suggesting a late O-type star. The stronger He\,{\sc ii} 4542\,\AA\ compared to Si\,{\sc iii} 4553\,\AA\ and stronger C\,{\sc iii} 4647/50/51\,\AA\ compared to He\,{\sc ii} 4686\,\AA\ are associated with spectral type O9.7. Due to the strength of Si\,{\sc ii} 4116\,\AA\ compared to He\,{\sc i} 4121\,\AA, this star is a dwarf star with spectral type O9.7 V. 

\subsection*{LS4943}
Similarly to LS4941, the ISIS spectrum of LS4943 shows He\,{\sc i} and He\,{\sc ii} lines and a strong He\,{\sc ii} 4686\,\AA\ compared to He\,{\sc i} 4713\,\AA, which is suggesting a late O-type classification, in line with the O9 V photometry classification of \cite{Povich2009}. By the same reasoning as for LS4941, this star is classified as O9.7 V. 

\subsection*{LS4972}
\cite{Povich2009} labelled this star as B2 V, but the ISIS spectrum of LS4972 reveals a SB2 nature with similarly strong He\,{\sc i} lines in both components. The lack of He\,{\sc ii} lines and strong He\,{\sc i} 4471\,\AA\ compared to the Mg\,{\sc ii} 4481\,\AA, suggests B1-B4 for both of the SB2 components. Since He\,{\sc i} 4121\,\AA\ is similar in strength to He\,{\sc i} 4144\,\AA, the stars are of B1-B2 spectral type. As the stars show a slight difference in strength of O\,{\sc i} 4070/76\,\AA, He\,{\sc i} 4144\,\AA\ and He\,{\sc i} 4387\,\AA, the classification is B1 + B2. Both stars show a relative lack of metal lines, yielding B1 V + B2 V. 

\subsection*{B181}
\cite{Povich2009} classified B181 (or CEN61) as O9V based on photometry and \cite{Hanson1997} as O9-B2 based on its K-band spectrum. The X-Shooter spectrum of B181 reveals several He\,{\sc i} lines and a few He\,{\sc ii} lines. The relative strengths of these lines point to a late O or early B-type star. The relative strength of the He\,{\sc ii} 4686\,\AA, C\,{\sc iii} 4647/50/51\,\AA\ and He\,{\sc i} 4713\,\AA\ and the weak He\,{\sc ii} 4200\,\AA, of similar strength to C\,{\sc iii} 4187\,\AA, point to a spectral type O9.7. The strength of the Si\,{\sc iv} 4116\,\AA\ compared to its neighbouring He\,{\sc i} 4121\,\AA\ points to luminosity class III. Therefore, the spectral type of B181 is O9.7 III.

\subsection*{BD-16 4822}
The star has been tentatively classified as O6.5 V by \cite{Povich2009} based on photometry, but a lack of He\,{\sc ii} lines in the ISIS spectrum points towards an early B classification rather than an O-type star. The relative strength between He\,{\sc i} 4471\,\AA\ and Mg\,{\sc ii} 4481\,\AA\ is indicative for spectral type B1-B3. Because of the presence of C\,{\sc iii} 4068/70\,\AA\ and the strength of He\,{\sc i} 4121\,\AA\ compared to Mg\,{\sc ii} 4481\,\AA\ this is specified to B2.5. The luminosity class of BD-164822 is I-III due to the relatively narrow shape of the hydrogen lines and the presence of N\,{\sc ii} 3995\,\AA. More specifically, the relative strength of Si\,{\sc ii} 4128-4130\,\AA\ points to luminosity class II, such that we settle on B2.5 II.

\section{Members}
\label{sec:appendix_members}
\onecolumn
\begin{landscape}
\topcaption{Members of NGC\,6618.}
\begin{supertabular}{llllllllll}
\hline
\hline
\noalign{\smallskip}source\_id & l & b & parallax & pml & pmb & Radial velocity & G & G$_{\rm{Bp}}$ & G$_{\rm{Rp}}$ \\
\noalign{\smallskip}- & deg & deg & mas & mas yr$^{-1}$ & mas yr$^{-1}$ & km s$^{-1}$ & mag & mag & mag \\
\hline
\noalign{\smallskip}4097809163052250624 & 15.0328 & -0.6955 & 0.653 $\pm$ 0.07 & -1.657 $\pm$ 0.074 & -1.387 $\pm$ 0.069 & - & 15.016 & 16.809 & 13.758 \\
\noalign{\smallskip}4097809163052248832 & 15.0339 & -0.6983 & 0.662 $\pm$ 0.095 & -1.257 $\pm$ 0.092 & -0.724 $\pm$ 0.103 & - & 16.853 & 18.228 & 15.407 \\
\noalign{\smallskip}4097809437930173440 & 15.0409 & -0.6756 & 0.565 $\pm$ 0.07 & -1.912 $\pm$ 0.065 & -0.409 $\pm$ 0.067 & - & 16.037 & 16.731 & 14.51 \\
\noalign{\smallskip}4097815244725944064 & 15.0433 & -0.695 & 0.594 $\pm$ 0.053 & -1.311 $\pm$ 0.059 & -0.72 $\pm$ 0.054 & - & 15.481 & 17.945 & 14.148 \\
\noalign{\smallskip}4097815309143033088 & 15.0443 & -0.6775 & 0.584 $\pm$ 0.074 & -1.643 $\pm$ 0.068 & -0.735 $\pm$ 0.07 & - & 14.645 & 15.748 & 13.454 \\
\noalign{\smallskip}4097815244725951744 & 15.0468 & -0.6846 & 0.63 $\pm$ 0.06 & -1.394 $\pm$ 0.057 & -1.112 $\pm$ 0.061 & - & 15.318 & 16.635 & 14.036 \\
\noalign{\smallskip}4097815038567510144 & 15.047 & -0.7024 & 0.572 $\pm$ 0.039 & -1.457 $\pm$ 0.038 & -1.496 $\pm$ 0.043 & - & 14.826 & 15.768 & 13.783 \\
\noalign{\smallskip}4097815244725949056 & 15.0487 & -0.6893 & 0.517 $\pm$ 0.037 & -1.54 $\pm$ 0.041 & -0.909 $\pm$ 0.038 & - & 14.848 & 16.295 & 13.644 \\
\noalign{\smallskip}4097815244725953280 & 15.0494 & -0.6837 & 0.751 $\pm$ 0.067 & -1.004 $\pm$ 0.065 & -0.927 $\pm$ 0.068 & 89.993 $\pm$ 9.195 & 15.381 & 16.782 & 14.139 \\
\noalign{\smallskip}4098003055053986816 & 15.0499 & -0.6474 & 0.568 $\pm$ 0.108 & -0.78 $\pm$ 0.111 & -2.008 $\pm$ 0.104 & - & 17.277 & 18.006 & 16.255 \\
\noalign{\smallskip}4098003055053984256 & 15.0514 & -0.6524 & 0.678 $\pm$ 0.091 & -0.756 $\pm$ 0.085 & -0.258 $\pm$ 0.088 & -38.879 $\pm$ 7.867 & 16.06 & 19.373 & 14.545 \\
\noalign{\smallskip}4097815313445442816 & 15.0518 & -0.6689 & 0.546 $\pm$ 0.029 & -1.127 $\pm$ 0.027 & -0.844 $\pm$ 0.028 & - & 14.07 & 15.446 & 12.871 \\
\noalign{\smallskip}4097815244725951488 & 15.0527 & -0.6871 & 0.628 $\pm$ 0.052 & -0.91 $\pm$ 0.049 & -0.911 $\pm$ 0.054 & - & 15.974 & 17.983 & 14.725 \\
\noalign{\smallskip}4097815244725950720 & 15.0527 & -0.688 & 0.537 $\pm$ 0.045 & -1.259 $\pm$ 0.049 & -0.52 $\pm$ 0.046 & - & 15.455 & 17.59 & 14.173 \\
\noalign{\smallskip}4097815004207762944 & 15.0532 & -0.7186 & 0.663 $\pm$ 0.067 & -0.814 $\pm$ 0.057 & -0.653 $\pm$ 0.059 & - & 16.247 & 16.469 & 15.211 \\
\noalign{\smallskip}4097815072927248384 & 15.0534 & -0.7045 & 0.555 $\pm$ 0.069 & -1.103 $\pm$ 0.075 & -0.969 $\pm$ 0.076 & - & 15.566 & 18.776 & 14.205 \\
\noalign{\smallskip}4098002986334497664 & 15.0542 & -0.6688 & 0.574 $\pm$ 0.07 & -1.852 $\pm$ 0.065 & -1.682 $\pm$ 0.068 & - & 15.59 & 16.947 & 14.316 \\
\noalign{\smallskip}4097815343502780544 & 15.0551 & -0.6845 & 0.441 $\pm$ 0.111 & -1.695 $\pm$ 0.114 & -0.808 $\pm$ 0.115 & - & 17.012 & 19.447 & 15.729 \\
\noalign{\smallskip}4098002986334501504 & 15.0554 & -0.6622 & 0.666 $\pm$ 0.099 & -1.582 $\pm$ 0.092 & -1.555 $\pm$ 0.096 & - & 16.593 & 18.194 & 15.202 \\
\noalign{\smallskip}4097815279085687808 & 15.0554 & -0.6915 & 0.501 $\pm$ 0.099 & -0.969 $\pm$ 0.098 & -1.369 $\pm$ 0.104 & 133.901 $\pm$ 6.741 & 15.836 & 20.172 & 14.256 \\
\noalign{\smallskip}4097815313445439872 & 15.0555 & -0.6738 & 0.556 $\pm$ 0.067 & -1.159 $\pm$ 0.062 & -1.236 $\pm$ 0.066 & - & 15.881 & 17.248 & 14.682 \\
\noalign{\smallskip}4097815274778072832 & 15.0565 & -0.6885 & 0.472 $\pm$ 0.049 & -1.493 $\pm$ 0.052 & -1.737 $\pm$ 0.052 & - & 12.153 & 14.173 & 10.724 \\
\noalign{\smallskip}4097815313445441664 & 15.0566 & -0.6718 & 0.595 $\pm$ 0.028 & -1.262 $\pm$ 0.028 & -0.525 $\pm$ 0.027 & - & 14.51 & 15.575 & 13.388 \\
\noalign{\smallskip}4097815347805179904 & 15.0576 & -0.6722 & 0.66 $\pm$ 0.059 & -1.897 $\pm$ 0.055 & -1.523 $\pm$ 0.058 & 32.397 $\pm$ 11.323 & 15.437 & 16.615 & 14.18 \\
\noalign{\smallskip}4097815279085690240 & 15.0595 & -0.6884 & 0.564 $\pm$ 0.055 & -1.83 $\pm$ 0.057 & -0.684 $\pm$ 0.059 & - & 15.109 & 17.267 & 13.691 \\
\noalign{\smallskip}4097815347805176704 & 15.0595 & -0.6758 & 0.65 $\pm$ 0.082 & -1.619 $\pm$ 0.075 & -0.762 $\pm$ 0.082 & - & 16.281 & 17.729 & 14.948 \\
\noalign{\smallskip}4098003020694235904 & 15.0603 & -0.6713 & 0.68 $\pm$ 0.084 & -1.378 $\pm$ 0.078 & -1.472 $\pm$ 0.085 & - & 16.133 & 17.613 & 14.845 \\
\noalign{\smallskip}4097815171704112256 & 15.0635 & -0.707 & 0.566 $\pm$ 0.083 & -0.547 $\pm$ 0.082 & -0.717 $\pm$ 0.134 & - & 16.576 & 17.225 & 15.096 \\
\noalign{\smallskip}4098003020694241024 & 15.0638 & -0.6638 & 0.721 $\pm$ 0.068 & -1.705 $\pm$ 0.066 & -1.36 $\pm$ 0.068 & - & 16.239 & 18.234 & 14.907 \\
\noalign{\smallskip}4097815382164896896 & 15.0644 & -0.7032 & 0.669 $\pm$ 0.094 & -1.245 $\pm$ 0.098 & -1.248 $\pm$ 0.103 & - & 16.766 & 18.61 & 15.395 \\
\hline
\hline
\end{supertabular}

\clearpage

\renewcommand\thetable{B.1}
\topcaption{Continued.}
\begin{supertabular}{l l l l l l l l l l l l l}
\hline
\hline
\noalign{\smallskip}source\_id & l & b & parallax & pml & pmb & radial\_velocity & G & G$_{\rm{Bp}}$ & G$_{\rm{Rp}}$ \\
\noalign{\smallskip}- & deg & deg & mas & mas yr$^{-1}$ & mas yr$^{-1}$ & km s$^{-1}$ & mag & mag & mag \\
\hline
\noalign{\smallskip}4097815171704114560 & 15.0646 & -0.7101 & 0.617 $\pm$ 0.018 & -1.285 $\pm$ 0.018 & -1.337 $\pm$ 0.019 & - & 12.88 & 13.372 & 12.178 \\
\noalign{\smallskip}4097815176006461440 & 15.0659 & -0.7123 & 0.608 $\pm$ 0.023 & -1.43 $\pm$ 0.023 & -1.663 $\pm$ 0.024 & - & 14.161 & 14.746 & 13.401 \\
\noalign{\smallskip}4097815347805170944 & 15.0673 & -0.6869 & 0.611 $\pm$ 0.023 & -1.299 $\pm$ 0.022 & -0.608 $\pm$ 0.025 & - & 13.809 & 15.495 & 12.555 \\
\noalign{\smallskip}4098003020694241920 & 15.0699 & -0.6646 & 0.669 $\pm$ 0.07 & -1.674 $\pm$ 0.058 & -1.251 $\pm$ 0.06 & - & 16.236 & 18.289 & 14.943 \\
\noalign{\smallskip}4097815382164903040 & 15.0701 & -0.6956 & 0.598 $\pm$ 0.054 & -1.148 $\pm$ 0.058 & -1.231 $\pm$ 0.088 & - & 15.753 & 17.678 & 14.418 \\
\noalign{\smallskip}4097815450884394112 & 15.0716 & -0.6783 & 0.734 $\pm$ 0.093 & -1.97 $\pm$ 0.086 & -1.292 $\pm$ 0.091 & - & 16.915 & 19.835 & 15.594 \\
\noalign{\smallskip}4097815382164899840 & 15.0732 & -0.7004 & 0.55 $\pm$ 0.021 & -1.142 $\pm$ 0.023 & -1.444 $\pm$ 0.022 & - & 10.512 & 11.354 & 9.584 \\
\noalign{\smallskip}4097814381430170496 & 15.0733 & -0.7429 & 0.574 $\pm$ 0.097 & -2.138 $\pm$ 0.11 & -0.243 $\pm$ 0.097 & - & 17.139 & 18.198 & 16.315 \\
\noalign{\smallskip}4098003123773451520 & 15.0734 & -0.6739 & 0.526 $\pm$ 0.045 & -1.493 $\pm$ 0.042 & -1.142 $\pm$ 0.045 & - & 15.492 & 17.281 & 14.238 \\
\noalign{\smallskip}4098003085110461312 & 15.0758 & -0.6532 & 0.59 $\pm$ 0.031 & -1.392 $\pm$ 0.029 & -1.193 $\pm$ 0.03 & - & 14.183 & 15.927 & 12.917 \\
\noalign{\smallskip}4097815450884392704 & 15.079 & -0.6819 & 0.662 $\pm$ 0.021 & -1.795 $\pm$ 0.02 & -0.35 $\pm$ 0.02 & - & 12.619 & 12.805 & 12.299 \\
\noalign{\smallskip}4097815210366202240 & 15.0811 & -0.7131 & 0.571 $\pm$ 0.043 & -1.446 $\pm$ 0.052 & -1.003 $\pm$ 0.05 & -6.883 $\pm$ 6.049 & 15.246 & 16.118 & 14.496 \\
\hline
\hline
\end{supertabular}

\end{landscape}
\twocolumn

\end{appendix}

\end{document}